# Triple-Resonance Spectroscopy Using a Cavity-Enhanced Frequency Comb Probe

**Qinxue Nie[1], Vinicius Silva de Oliveira[1], Adrian Hjältén[1], Isak Silander[1], Kevin K. Lehmann[2], and Aleksandra Foltynowicz[1,*]**

[1] Department of Physics, Umeå University, 901 87 Umeå, Sweden

[2] Departments of Chemistry & Physics, University of Virginia, Charlottesville, VA 22904, USA

[*]Corresponding author: aleksandra.foltynowicz@umu.se

**Abstract:** Accurate experimentally verified models of molecular hot-band transitions are essential for interpreting high-temperature spectra in environments ranging from exoplanetary atmospheres to combustion systems. However, line-resolved measurements of infrared hot-band transitions reaching highly excited vibrational states above 10000 $cm^{-1}$ are missing because these transitions are too weak to observe at room temperature and the spectra are congested at elevated temperatures. Here, we introduce a nonlinear spectroscopic approach that enables simultaneous measurement of individual infrared hot-band transitions between five vibrational bands up to 12000 $cm^{-1}$ over a broad spectral range with Doppler-free resolution and sub-MHz frequency accuracy ($10^{-10}$ relative line position accuracy). The method is based on an all-optical triple-resonance (AOTR) scheme that combines stepwise mid-infrared pumping using an optical-frequency-comb-stabilized, double-seeded continuous-wave optical parametric oscillator with broadband highly sensitive near-infrared probing using a cavity-enhanced optical frequency comb. As a proof of principle, we measure transitions between high polyads (*P*) of methane – groups of strongly interacting, near-degenerate vibrational energy states arising from couplings between the C-H stretching and bending modes. In the probe spectra, we simultaneously resolve Doppler-free $P4 \leftarrow P0$, $P6 \leftarrow P2$ and $P8 \leftarrow P4$ transitions, reaching the poorly understood polyad *P*8 near 12000 $cm^{-1}$ and providing the first set of 105 experimentally observed lines in the $P8 \leftarrow P4$ spectral region, of which 39 lines are assigned, reaching 24 unique final states. Comb-based AOTR spectroscopy opens a new route for broadband exploration of highly excited molecular states, delivering extensive high-accuracy spectroscopic data needed to refine molecular models and improve predictions of high-temperature spectra in astrophysical and combustion environments.

## 1. Introduction

Recently, the James Webb Space Telescope (JWST) provided robust evidence of the existence of methane ($CH_4$) in the atmosphere of WASP-80b, a Jupiter-like exoplanet with a temperature of 825 K, hotter than any planet in the Solar System[1]. Yet, in the context of exoplanets, WASP-80b is considered only 'warm', because extreme exoplanetary environments can reach much higher temperatures, approaching 5000 K[2]. At such high temperatures, dense hot-band manifolds produce highly congested spectra that are difficult to analyze, because accurate line lists of transitions reaching the relevant highly-excited states are missing for many molecules in the literature and spectroscopic databases[3–5]. This is because high-temperature laboratory spectra do not provide line-resolved spectral information[6,7], and existing nonlinear[8–11] and non-equilibrium techniques[12] do not reach states above 10000 $cm^{-1}$ with a sufficient combination of broad spectral bandwidth and high resolution. The poor characterization of these hot-band transitions limits the accurate analysis of spectral data from the extremely hot exoplanets[13,14]. To assess and improve the accuracy of spectral models used to interpret such spectra, high-precision measurements of excited molecular states are needed[15].

High-temperature spectra of methane pose a particular challenge for modeling because of its remarkably complex vibrational energy level structure. The nearly-resonant fundamental vibrational modes and their overtones are strongly coupled through Fermi, Darling–Dennison and Coriolis interactions, forming groups of states called polyads with rapidly increasing state density[16]. The *P*n polyad is the group of states for which n = $2n_1 + n_2 + 2n_3 + n_4$, where $n_i$ is the number of quanta in vibrational normal mode i. Each polyad is centered roughly at 1500×n cm$^{-1}$. The most complete assigned high-temperature line list of methane in the ExoMol database[5] contains states up to 18000 cm$^{-1}$ (*P*13), but only a small fraction (0.25%) of levels has been experimentally verified up to 9900 cm$^{-1}$ (*P*6 and *P*7)[17], and its accuracy decreases with increasing polyad number. The energy levels above 10000 cm$^{-1}$ have been reached mainly by measuring overtone transitions from the ground state at room or low temperatures, using Fourier transform infrared spectroscopy[18,19], intracavity laser absorption spectroscopy[20–22], laser-induced grating spectroscopy[23,24], intracavity photoacoustic spectroscopy[25], and cavity ringdown spectroscopy[19]. All these studies found that even the strongest bands in the *P*8 polyad are highly congested: isolated lines are mostly absent and a substantial unresolved background is present; moreover, the limited possibility of applying lower-state combination-difference relations hinders reliable validation of rovibrational assignments[19]. Consequently, spectral congestion in high polyads prevents spectral assignment of ground state transitions, leaving regions of *P*8 and above largely uncharacterized. This limits the accuracy of high-temperature spectral modeling and motivates the need for new spectroscopic approaches.

Optical-optical double-resonance (OODR) spectroscopy using narrow-linewidth continuous-wave (CW) lasers is a powerful tool for probing hot-band transitions from selectively populated intermediate states, yielding resolved Doppler-free absorption features[9,26–28]. Double- and triple-resonance spectroscopies based on pulsed lasers have been used to reach very highly excited molecular states (>15000 cm$^{-1}$), though with a resolution insufficient to observe Doppler-free features[28–30]. Doppler-free-resolution OODR spectroscopy reaching the spectral region near 12000 cm$^{-1}$ has been reported for $CO_2$, using two cavity-enhanced CW near-infrared (NIR) lasers as pump and probe[9]. However, the narrow spectral range of CW lasers inherently limits the accessible spectral range, and coupling both the pump and probe to the same cavity prevents independent control of their frequencies. This makes CW-OODR impractical for searching hot-band transitions whose frequencies are not known with sub-GHz accuracy. Using an optical frequency comb as the probe in OODR opens a highly efficient way to search transitions over a broad spectral range with high frequency accuracy and precision[11]. However, to date, comb-based OODR spectroscopy has reached excited states of methane only up to the *P*6 polyad[11,31–34], i.e., up to the 9400 cm$^{-1}$ range.

Here, we introduce all-optical triple-resonance (AOTR) spectroscopy using a cavity-enhanced comb probe to investigate an unprecedentedly broad range of highly excited molecular states from ~6000 cm$^{-1}$ up to the unexplored region near 12000 cm$^{-1}$ with Doppler-free precision and high absorption sensitivity. As a demonstration of principle, we simultaneously detect methane transitions in the $P4 \leftarrow P0$, $P6 \leftarrow P2$ and $P8 \leftarrow P4$ ranges, up to the pentacontakaipentad polyad ($P8$) near 12000 cm$^{-1}$. We report the first set of high-precision spectral data in the $P8 \leftarrow P4$ region, with 39 out of 105 probed transitions in this region assigned based on combination-difference measurements and ExoMol predictions.

The AOTR scheme creatively employs stepwise mid-infrared (MIR) pumping to populate two intermediate rovibrational levels. The two MIR pump fields used in this work are two idlers simultaneously generated by a CW singly-resonant optical parametric oscillator (CW-OPO) employing dual seeding of the pump amplifier, hereafter termed a double-seeded CW-OPO. While double-seeded CW-OPOs using either two CW seeds[35–38] or dual-comb seeds[39,40] have been reported previously, their use in high-precision spectroscopy has been limited by the lack of comb-referenced stabilization schemes. Here, we demonstrate comb-referenced frequency stabilization of both idlers for the first time, enabling applications of double-seeded CW-OPOs in high-precision spectroscopy. The comb-based AOTR platform opens a way to investigate highly excited molecular states to address gaps in spectral databases and provides high-accuracy data for refining models of high-temperature spectra.

## 2. Results

### *2.1 Principle*

The general energy level structure of the AOTR scheme designed to reach the $P8$ polyad of methane around 12000 cm$^{-1}$ is depicted in Fig. 1a.

Methane molecules are excited to a selected assigned rovibrational level in the *P*4 polyad via two-step pumping using two CW MIR fields around 3000 cm$^{-1}$: the *P*2←*P*0 transition pumped by pump1, followed by the *P*4←*P*2 transition pumped by pump2. The narrow-linewidth pumping is velocity selective[41], and the probe transitions are free of Doppler broadening. The *P*8 and *P*6 polyads are reached by ladder-type *P*8←*P*4 and *P*6←*P*2 transitions probed by the NIR comb centered around 6000 cm$^{-1}$. These transitions appear as Doppler-free peaks in the spectrum, and the *P*8←*P*4 transitions are narrower than the *P*6←*P*2 transitions because of the more restrictive velocity selection by two pumps compared to one. Moreover, V-type *P*4←*P*0 transitions appear in the same spectrum as Doppler-free dips in Doppler-broadened lines that share the depleted lower state with the *P*2←*P*0 transition excited by pump1. Therefore, the AOTR is a unique scheme that links five polyads and simultaneously detects three types of Doppler-free transitions in the probe spectrum, allowing retrieval of high-accuracy spectral data in the *P*8, *P*6, and *P*4 polyads over a broad spectral range within a single measurement.

Since the best current theoretical predictions[5] of *P*8←*P*4 transitions are not sufficiently accurate for unambiguous line assignment based on closest frequency and intensity match, we use combination-difference measurements to enable experimental assignment of the rotational quantum numbers of the final states, as shown in Fig. 1b. Starting from a common ground state level with a rotational quantum number $J = 2$ and symmetry $C = F_2$, three levels in the $2\nu_3$ state in the $P4$ polyad are populated through different pump transition pairs: $2F_2$ via Q(2, $F_2$) and Q(2, $F_1$) transitions, $3F_2$ via Q(2, $F_2$) and R(2, $F_1$) transitions, and $4F_2$ via R(2, $F_2$) and R(3, $F_1$) transitions. Depending on how many $P8 \leftarrow P4$ transitions (1, 2 or 3) from these different $P4$ states reach a given upper state (which are all of $F_1$ symmetry), its rotational quantum number $J$ can be unambiguously assigned according to the selection rules $\Delta J = 0$ or $\pm 1$.

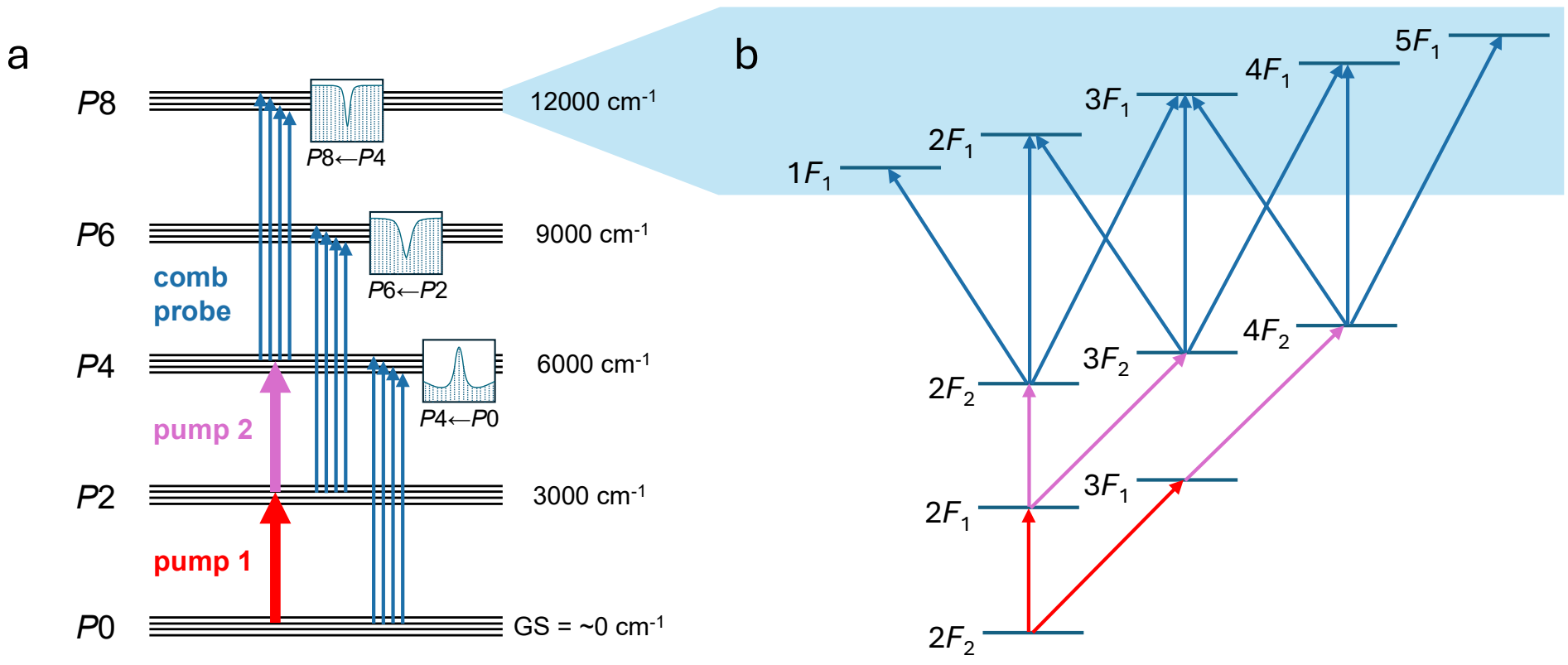

**Fig. 1 Energy levels involved in AOTR spectroscopy. a** Schematic of the vibrational polyads of methane addressed by the two CW MIR pump fields and the NIR comb probe. $P8 \leftarrow P4$ and $P6 \leftarrow P2$ correspond to two types of ladder-type transitions, while $P4 \leftarrow P0$ represents V-type transitions, all of which are simultaneously detected using the comb probe. GS: ground state. $P0$, $P2$, $P4$, $P6$ and $P8$ are vibrational polyad numbers. **b** A detailed view of rotational sublevels addressed by the pumps and the comb probe showing how combination differences can be used to assign the rotational quantum number of the final states in the $P8$ polyad.

### *2.2 Experimental setup and procedures*

Figure 2 shows the experimental setup of AOTR spectroscopy. The two pump fields, pump1 and pump2, operating at wavelengths near 3.31 μm, are two idlers generated by double-seeding a CW-OPO using two narrow-linewidth lasers emitting near 1.06 μm (see Supplementary Note 1). Both idler frequencies are stabilized at the frequencies of the respective pump transitions by a phase-lock to an MIR comb via feedback to the seed lasers. Since both idlers are emitted from the same CW-OPO cavity, they are inherently spatially overlapped, enabling efficient simultaneous interaction with the molecules.

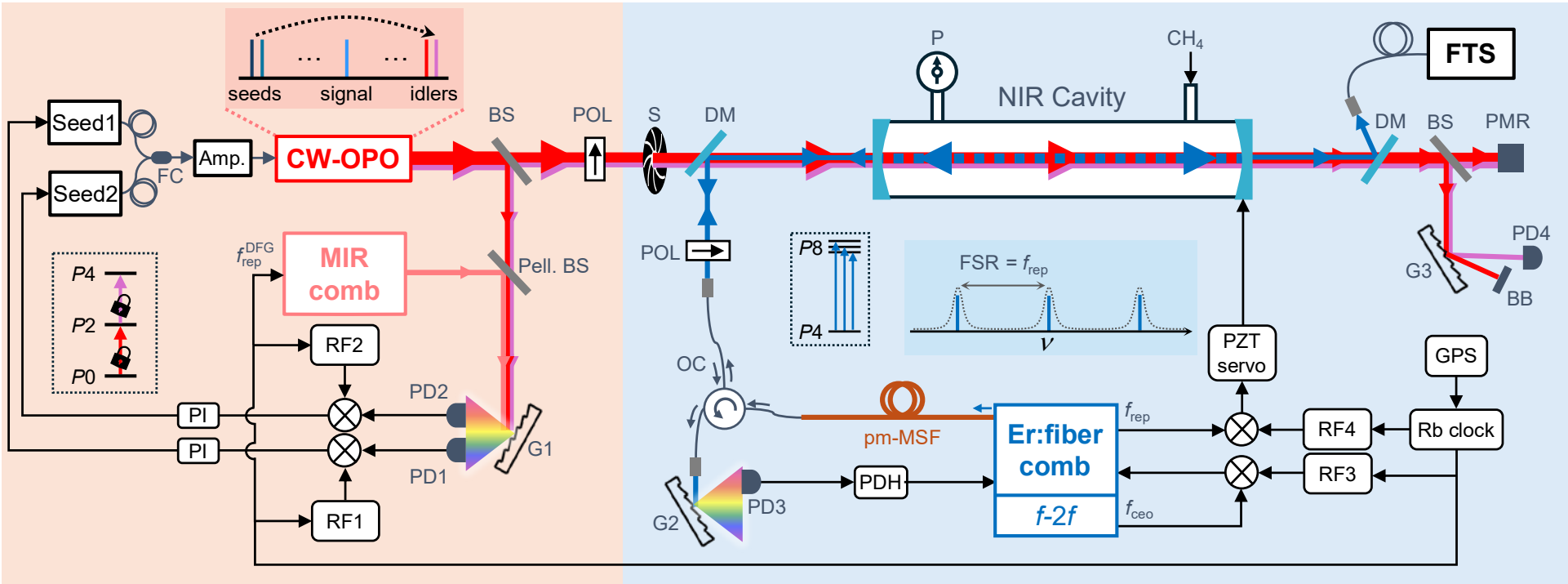

**Fig. 2 Experimental setup.** Seed: seed laser. FC: fiber coupler. Amp.: optical amplifier. CW-OPO: continuous-wave optical parametric oscillator. BS: beamsplitter. Pell. BS: pellicle beamsplitter. MIR comb: a mid-infrared frequency comb produced by difference frequency generation. HWP: half-wave plate. S: optical shutter. DM: dichroic mirror. POL: polarizer. PZT: piezoelectric transducer. P: pressure gauge. NIR cavity: near-infrared cavity. FTS: Fourier transform spectrometer. PMR: power meter. PI: proportional-integral controller. RF: radio frequency source. PD: photodiode. G: grating. OC: optical circulator. pm-MSF: polarization-maintaining microstructured silica fiber. *f*-2*f*: an *f*-2*f* interferometer. BB: beam block.

An NIR comb centered at 1.68 μm with a repetition rate ($f_{rep}$) of ~250 MHz is employed to probe transitions over a spectral range of ~7 THz. A 60-cm-long NIR optical cavity is used to extend the probe-molecule interaction length. The two idlers and the NIR comb probe are combined at the cavity input and separated at the cavity output by dichroic mirrors. The comb probe is mode-matched to the cavity's $TEM_{00}$ mode with a beam waist of 0.5 mm and a Rayleigh range of 45.8 cm. Both idlers are also focused at the center of the cavity via a common optical path with Rayleigh ranges matched to that of the comb beam to maximize pump-probe spatial overlap[34]. The cavity mirrors are antireflection-coated in the idler spectral range, providing ~95% single-pass transmission. The two pumps and the probe are linearly polarized and configured with parallel polarization alignment.

Stable comb-cavity coupling is achieved using the Pound-Drever-Hall (PDH) locking technique[42]. The cavity at room temperature is filled with pure $CH_4$ at a pressure of around 60 mTorr for all measurements. Cavity-transmitted comb light, with a bandwidth of ~200 $cm^{-1}$, is coupled to a fast-scanning Fourier transform spectrometer (FTS) with auto-balanced detection[43] to acquire probe spectra. Comb-mode-limited resolution is achieved using the sub-nominal sampling-interleaving method[44,45]. Probe spectra are sampled with a 2-MHz sample point spacing by interleaving spectra acquired at different $f_{rep}$ values. The baseline, originating from the cavity-transmitted comb envelope and the Doppler-broadened absorption signals of methane overtone transitions from thermally populated ground-state levels, is removed by normalizing the spectrum acquired with pump excitations to the spectrum acquired with both pumps blocked by an optical shutter. More experimental details are provided in Methods.

**Table 1.** The assignments, wavenumbers and Einstein A coefficients of the pump transitions used in the three AOTR measurements, together with the pump power levels and spectral coverage of the comb probe. The line centers and uncertainties of the $2\nu_3 \leftarrow \nu_3$ pump transitions are from OODR measurements (Supplementary Note 2). The Einstein A coefficients and the assignments for the lower and upper states of the transitions are from ExoMol[5] and have the form ($P$ $J$ $C$ $N$), where $P$ is the polyad number, $J$ the total angular momentum quantum number, $C$ the rovibrational symmetry, and $N$ the counting number.

| Measurement | $\nu_3$ pump transition assignment, wavenumber, and Einstein A coefficient | Power of pump1 [mW] | $2\nu_3 \leftarrow \nu_3$ pump transition assignment, wavenumber and Einstein A coefficient | Power of pump2 [mW] | Probe spectral coverage [cm$^{-1}$] |
|---|---|---|---|---|---|
| $2F_2$ AOTR | Q(2, $F_2$)<br>(2 2 $F_1$ 8)←(0 2 $F_2$ 1)<br>3018.65020715(7)[46] cm$^{-1}$<br>25.7[5] s$^{-1}$ | 176 | Q(2, $F_1$)<br>(4 2 $F_2$ 80)←(2 2 $F_1$ 8)<br>2984.878457(2) cm$^{-1}$<br>15.3[5] s$^{-1}$ | 286 | 5850-6050 |
| $3F_2$ AOTR | | 118 | R(2, $F_1$)<br>(4 3 $F_2$ 106)←(2 2 $F_1$ 8)<br>3017.111842(2) cm$^{-1}$<br>7.0[5] s$^{-1}$ | 432 | 5840-6070 |
| $4F_2$ AOTR | R(2, $F_2$)<br>(2 3 $F_1$ 13)←(0 2 $F_2$ 1)<br>3048.15331811(8)[47] cm$^{-1}$<br>27.2[5] s$^{-1}$ | 305 | R(3, $F_1$)<br>(4 4 $F_2$ 140)←(2 3 $F_1$ 13)<br>3027.162657(1) cm$^{-1}$<br>26.9[5] s$^{-1}$ | 309 | 5820-6020 |

We measured three AOTR spectra using three combinations of pump1 and pump2 frequencies to reach different rotational sublevels in the $2\nu_3$ band in the $P$4 polyad, namely $2F_2$, $3F_2$ and $4F_2$. Assignments, wavenumbers and Einstein A coefficients of the pump transitions are listed in Table 1, together with the pump power levels incident on the gas sample and comb probe spectral coverage of the three measurements. The wavenumbers of the $\nu_3$ pump transitions, Q(2, $F_2$) and R(2, $F_2$), are known with an uncertainty of 2 kHz and 2.5 kHz from Okubo *et al.*[46] and Abe *et al.*[47], respectively. We used MIR-MIR OODR to determine the wavenumbers of the $2\nu_3 \leftarrow \nu_3$ Q(2, $F_1$), R(2, $F_1$) and R(3, $F_1$) pump transitions with 55 kHz, 62 kHz and 34 kHz uncertainties, respectively, as described in Supplementary Note 2. All Einstein A coefficients are taken from ExoMol[5].

### *2.3 New transitions detected by AOTR*

Figure 3a shows a narrow section of the interleaved and normalized 3$F_2$ AOTR spectrum (red) containing Doppler-free ladder-type probe transitions. To clearly distinguish the $P8 \leftarrow P4$ and $P6 \leftarrow P2$ transitions, a spectrum acquired with pump2 turned off and the same power of pump1 is plotted in the same figure (black). Two Doppler-free absorption features observed exclusively in the AOTR spectrum originate from two $P8 \leftarrow P4$ transitions from the $2\nu_3$ state excited by both pumps, while the feature present in both spectra is a $P6 \leftarrow P2$ transition from the $\nu_3$ state excited by pump1. We note that the frequencies of all three selected $2\nu_3 \leftarrow \nu_3$ pump transitions do not match any ground state transitions to the $P2$ polyad of methane and therefore cannot populate any $P2$ state and induce any $P6 \leftarrow P2$ transitions. The insets demonstrate that the $P8 \leftarrow P4$ transitions have a narrower linewidth than the $P6 \leftarrow P2$ transition. Furthermore, excitation of the $2\nu_3 \leftarrow \nu_3$ pump transition by pump2 depopulates the $\nu_3$ state in the $P2$ polyad, thereby reducing the intensity of the $P6 \leftarrow P2$ transitions. Figure 3b shows a V-type transition in the unnormalized AOTR spectrum, appearing as a Doppler-free dip at the center of a Doppler-broadened $2\nu_3$ Q(2, $F_2$) transition.

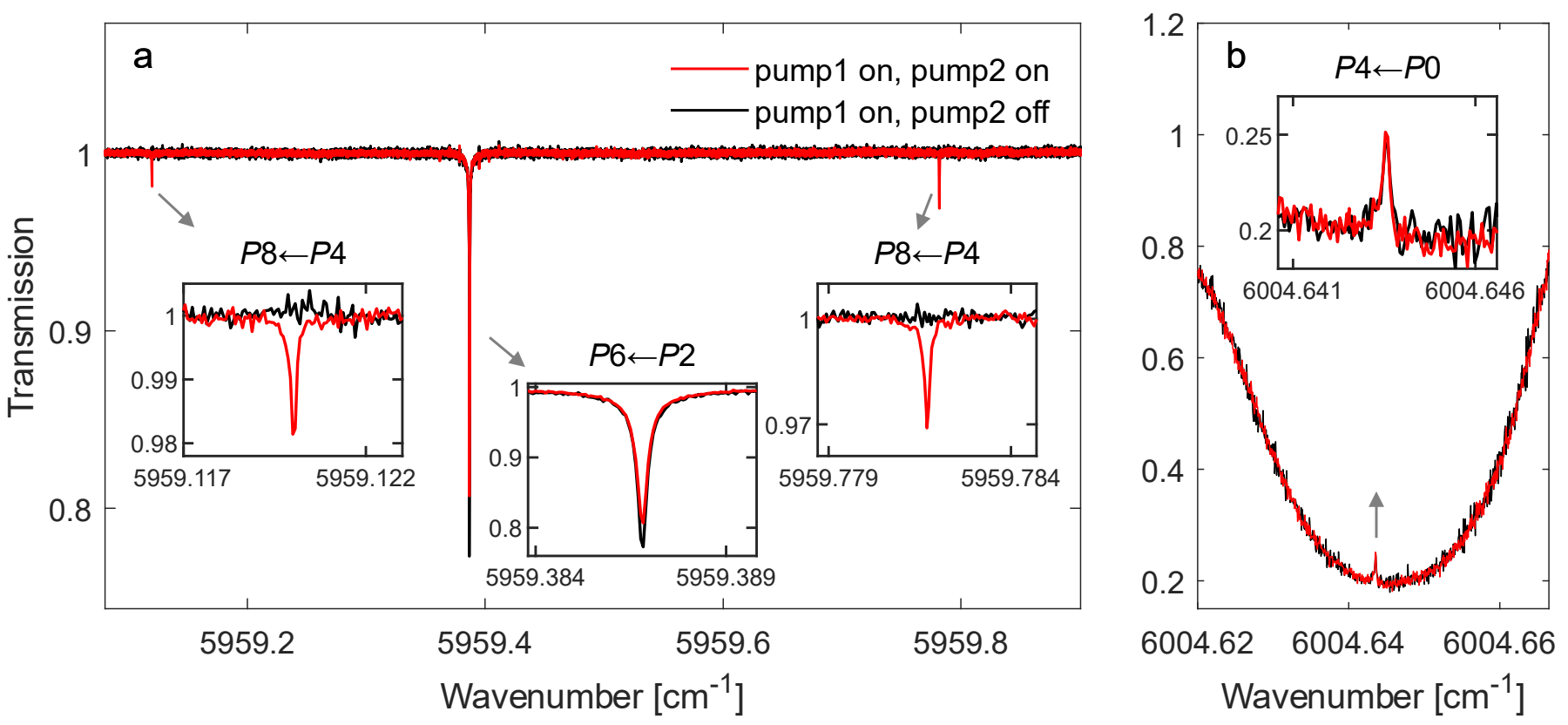

**Fig. 3. AOTR spectrum.** Narrow sections of $3F_2$ AOTR probe spectra measured under two pump conditions: pump1 on and pump2 on (red), pump1 on and pump2 off (black). **a** Normalized and interleaved AOTR spectrum with insets illustrating two $P8 \leftarrow P4$ transitions and a $P6 \leftarrow P2$ transition, respectively. **b** AOTR spectrum without normalization, with the inset illustrating a V-type $P4 \leftarrow P0$ transition.

We fit the cavity transmission function[48] to all Doppler-free probe transitions in the AOTR spectra to retrieve their line-center positions, integrated absorption coefficients and widths. The model includes a narrow Lorentzian component for the Doppler-free signal and a broader but weaker Gaussian component representing the elastic collision-induced Doppler-broadened background. Figure 4 shows a $P8 \leftarrow P4$ transition, a $P6 \leftarrow P2$ transition and a $P4 \leftarrow P0$ transition from the $3F_2$ AOTR measurement with their fits in the frequency range of ±100 MHz around the center, with residuals shown in the lower panels. Details of the fitting model and procedures are described in Supplementary Note 3, and fitting results for all lines are summarized in Tables S1-S3 in the Supplementary Note 4. The structured residual for the line shown in Fig. 4b indicates that the model does not fully capture all molecular velocity groups involved in the AOTR process and occurs only for strong ladder-type transitions with probe absorption exceeding 10%, accounting for approximately 5% of the measured ladder-type transitions. The asymmetry of the V-type transition in Fig. 4c is caused by the nonzero comb-cavity phase offset away from the locking point, and is included in the model.

The line positions of 105 $P8 \leftarrow P4$ transitions in the three AOTR measurements are retrieved with uncertainties ranging from 75 kHz to 1.1 MHz, those of the $P6 \leftarrow P2$ transitions – from 54 kHz to 1.5 MHz, and those of the V-type $P4 \leftarrow P0$ transitions – from 204 kHz to 1.1 MHz. In the $3F_2$ AOTR measurement, the integrated absorption coefficients of $P8 \leftarrow P4$ transitions are about one order of magnitude smaller than those of the $P6 \leftarrow P2$ transitions, primarily because of the lower population in the pumped $2\nu_3$ level in the $P4$ polyad relative to the $\nu_3$ level in the $P2$ polyad. The comparison of the two fits shown in Fig. 4b demonstrates that the absorbance of the $P6 \leftarrow P2$ transition is reduced by approximately 9% in the AOTR spectrum relative to the spectrum with pump2 off, indicating that pump2 depopulates the $\nu_3$ state via the $2\nu_3 \leftarrow \nu_3$ transition and correspondingly excites the same proportion of methane molecules into the $P4$ polyad.

The fitted half-width at half maximum (HWHM) of the $P8 \leftarrow P4$ transition shown in Fig. 4a is 2.26(7) MHz, notably narrower than that of the $P6 \leftarrow P2$ transition shown in Fig. 4b, which is 4.77(4) MHz. This trend is consistently observed across the three AOTR measurements. For the $2F_2$, $3F_2$, and $4F_2$ measurements, the mean HWHMs of the $P8 \leftarrow P4$ transitions are 2.5(4), 2.4(4), and 2.5(5) MHz, respectively (where the uncertainties are the standard deviations of the mean), approximately half of the corresponding mean HWHMs of the $P6 \leftarrow P2$ transitions, which are 5.4(5), 4.6(5), and 5.8(9) MHz, respectively. With the exception of the weakest $P6 \leftarrow P2$ transitions, for which the fit returns a narrow linewidth because of the low signal-to-noise ratio, the linewidth serves as an indicator to directly distinguish these two kinds of ladder-type transitions in the AOTR spectrum.

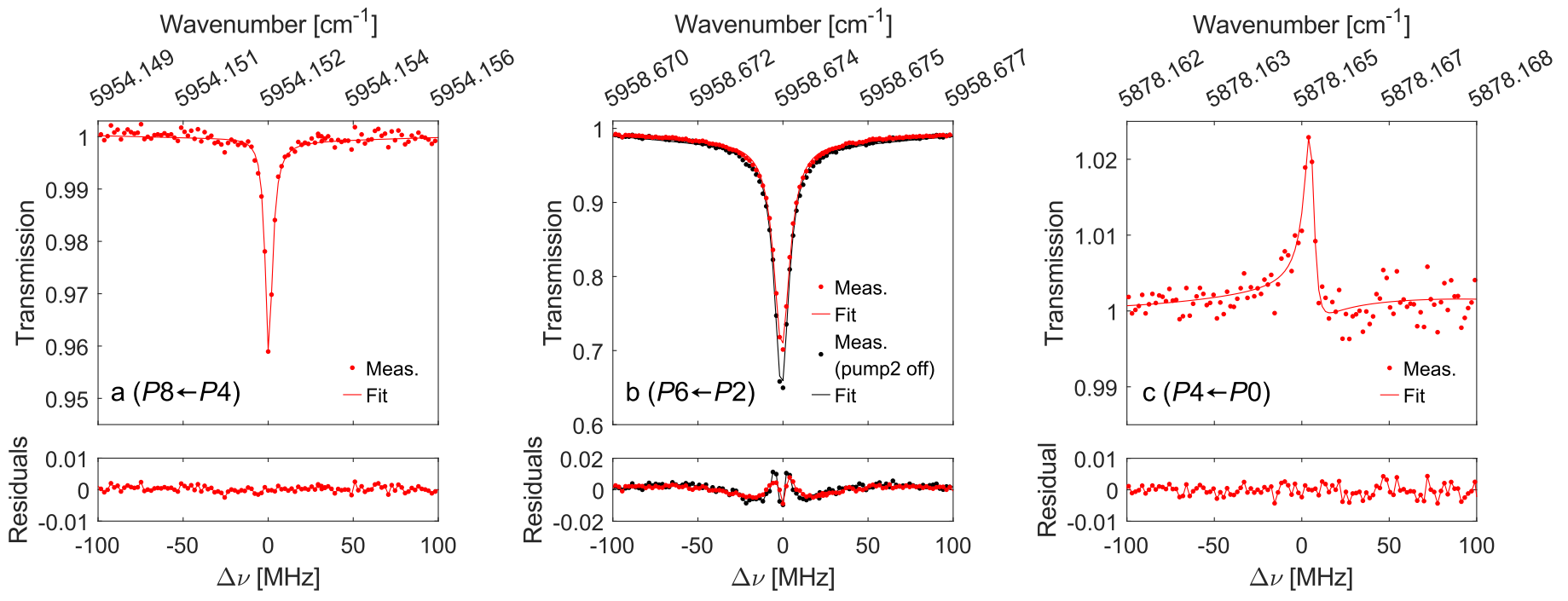

**Fig. 4. Three kinds of Doppler-free absorption signals in the AOTR spectrum with the fits. a** A $P8 \leftarrow P4$ transition. **b** A $P6 \leftarrow P2$ transition probed in the AOTR spectrum (red) and the spectrum with pump2 turned off (black). **c** A V-type $P4 \leftarrow P0$ transition. Upper panels: measured data (dots) with fits of the cavity transmission function (solid curves). Lower panels: residuals of the fits. $\Delta\nu$: frequency detuning from the line center of the transition.

### *2.4 Comparison to theoretical predictions*

In total, we found 105 $P8 \leftarrow P4$ transitions, 44 $P6 \leftarrow P2$ transitions, and 14 V-type $P4 \leftarrow P0$ transitions in the three AOTR spectra using a custom peak detection routine[33]. Figures 5-7 show all detected transitions as a function of the final state term value, compared to predictions from ExoMol[5], plotted inverted for clarity. The upper x axis in those figures shows the probe transitions wavenumbers. The ExoMol predictions include only the subset of transitions whose rovibrational lower level is the same as in the corresponding AOTR spectrum.

In the $P8 \leftarrow P4$ range, the ExoMol predictions have uncertainties ranging from 0.8 $cm^{-1}$ to 3.1 $cm^{-1}$, i.e., larger than the mean separation of the transitions from the pumped state in the $P4$ polyad. This precludes unambiguous assignments based on comparison of predicted and measured transition wavenumbers and intensities alone, as shown in Fig. 5a-c. Therefore, we used the combination difference approach to rotationally assign some of the transitions (see Fig. 1b). For each transition, we calculated the term value of the final state in the $P8$ polyad as the sum of the $2F_2$ ground state term value of 31.44238938(3) $cm^{-1}$ from de Oliveira *et al.*[49], the wavenumbers of the $\nu_3$ and the $2\nu_3 \leftarrow \nu_3$ pump transitions listed in Table 1, and the measured wavenumbers of the $P8 \leftarrow P4$ probe transitions in Table S2 in Supplementary Note 4. In the three AOTR measurements, we identified 3 final states in the $P8$ polyad reached by three different pump-probe combinations and 9 final states reached by two pump-probe combinations. These combination differences are illustrated in Fig. 5d, where the y-axis shows the deviation of each measured final state term value from the weighted mean of all measurements, with weights given by the inverse squared uncertainties. The error bars are the 1σ uncertainties of the individual final state term values, and for each upper state, the independently measured final state term values agree within their 3σ uncertainties. Table 2 shows the three pump-probe combinations that reach the same final state at 12021.356734(3) $cm^{-1}$, for which the rotational assignment thus must be $3F_1$. The only final state predicted in ExoMol consistent with this rotational assignment and the combination differences within 3.1 $cm^{-1}$ is the $4\nu_3$ state, for which the predicted intensities of the three transitions are also comparable to the measured relative line intensities. The mean shift of the three transitions is 2 $cm^{-1}$. In this way, we assigned all 27 transitions that form combination differences and found that the ExoMol transition wavenumbers exhibit different systematic positive offsets for different vibrational bands in the *P*8 polyad. We therefore applied these experimentally determined vibrational-band-dependent shifts to the ExoMol predictions for all bands that formed a combination difference. After applying these shifts and comparing to the experimental data, we tentatively assigned 12 additional transitions to four vibrational bands based on a wavenumber difference below 0.8 $cm^{-1}$ and comparable relative line intensities. All assignments for $P8 \leftarrow P4$ transitions probed in the three measurements are listed in Table S2 in the Supplementary Note 4, while a list of all assigned states in the $P8$ polyad is provided in Table S5. Fig. 5e plots the differences between the measured center wavenumbers and the unshifted ExoMol predictions for the assigned $P8 \leftarrow P4$ transitions, which have a mean value of -2.59 $cm^{-1}$ and standard deviation of 0.57 $cm^{-1}$, as summarized in Table 3. This disagreement is caused by the inaccuracy of the ExoMol predictions, because the uncertainties of our measurements are approximately five orders of magnitude smaller than those of the ExoMol predictions.

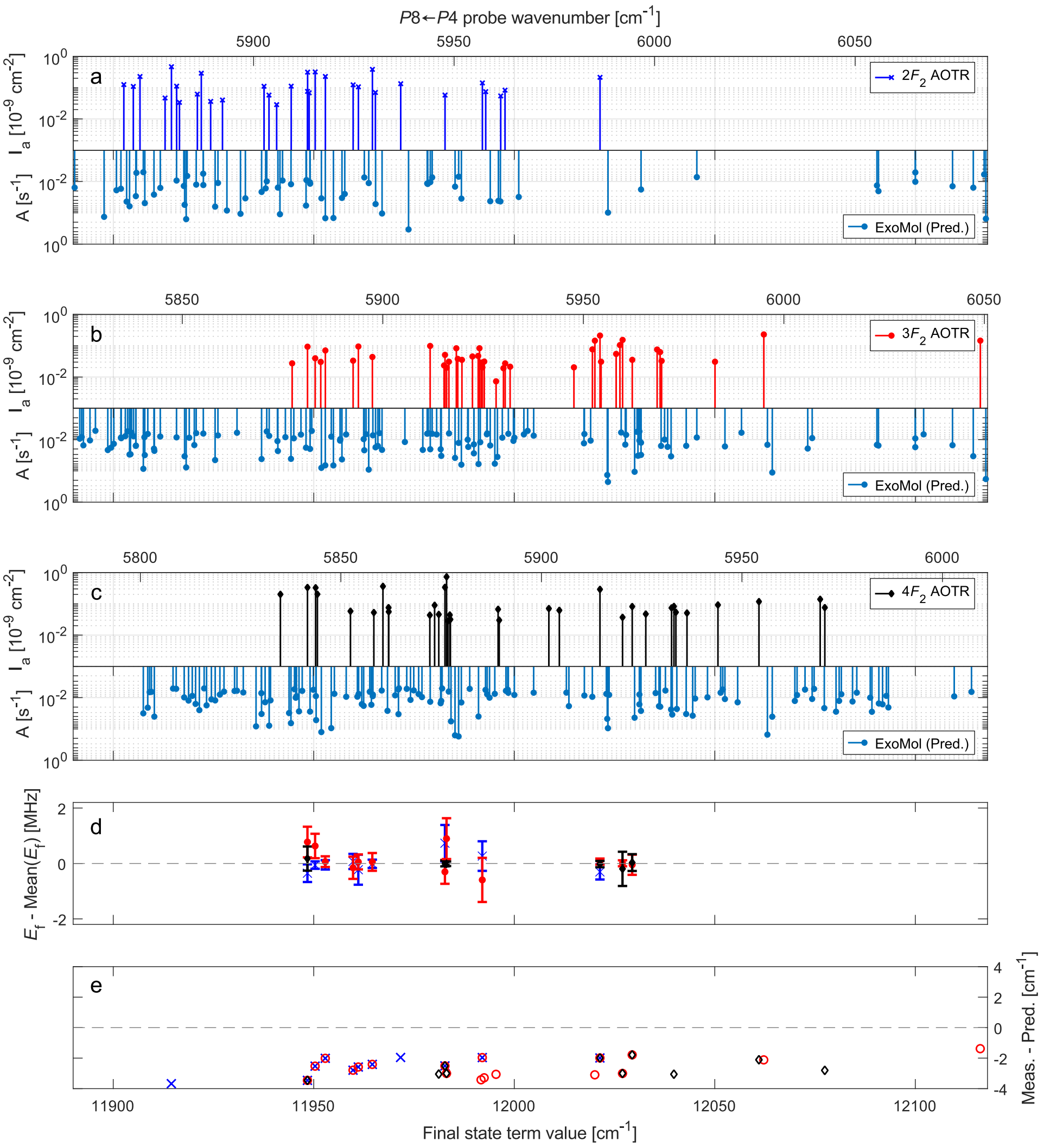

**Fig. 5. Measured $P8 \leftarrow P4$ transitions and comparison to ExoMol predictions.** The integrated absorption coefficients, $I_a$, of the ladder-type transitions in the $P8 \leftarrow P4$ range measured in **a** $2F_2$ AOTR, **b** $3F_2$ AOTR and **c** $4F_2$ AOTR are plotted in the upper panels, while the Einstein A coefficients, A, of the corresponding ExoMol predictions are shown in the bottom panels, inverted for clarity. The upper x axis shows the probe wavenumbers, and the bottom x axis, common in all plots, shows the final state term values reached by the transitions. **d** The difference between the individual final state term values, $E_f$, in the $P8$ polyad and their weighted mean for the transitions that form combination differences. **e** Differences between the measured center wavenumbers and the ExoMol predictions of the assigned $P8 \leftarrow P4$ transitions.

**Table 2.** Three pump-probe combinations that reach the same final state in the $P8$ polyad, which allows vibrational and rotational assignment of the final state ($n_1$ $n_2$ $n_3$ $n_4$ $C_v$) and ($P$ $J$ $C$ $N$), respectively, from ExoMol.

| Measurement | Pump transitions | $2\nu_3$ state term value in $P4$[5] [$cm^{-1}$] | Probe transition assignment | Probe transition wavenumber [$cm^{-1}$] | Final state term value in $P8$ [$cm^{-1}$] | Final state assignment[5] |
|---|---|---|---|---|---|---|
| $2F_2$ AOTR | Q(2, $F_2$) then Q(2, $F_1$) | 6034.971054(2) | R(2, $F_2$) | 5986.385673(9) | 12021.356726(9) | 0 0 4 0 $E$ (8 3 $F_1$ 2011) |
| $3F_2$ AOTR | Q(2, $F_2$) then R(2, $F_1$) | 6067.204439(2) | Q(3, $F_2$) | 5954.152300(3) | 12021.356739(4) | |
| $4F_2$ AOTR | R(2, $F_2$) then R(3, $F_1$) | 6106.758364(1) | P(4, $F_2$) | 5914.598372(4) | 12021.356736(4) | |

The measured $P6 \leftarrow P2$ transitions from the pumped $\nu_3$ $2F_1$ and $3F_1$ states in the $P2$ polyad are plotted in Fig. 6a and b, respectively. The agreement with ExoMol in this range is significantly better, because the predictions have been refined by experimental data[5]. The final state term values in the $P6$ polyad are calculated as the sum of the $2F_2$ ground state term value, the $\nu_3$ pump transition wavenumber, and the measured wavenumbers of the $P6 \leftarrow P2$ probe transitions. Since the $2F_2$ AOTR and $3F_2$ AOTR measurements used the same $\nu_3$ pump transition, we took the weighted mean of the probe transition wavenumbers observed in both spectra, using the inverse squared uncertainties as weights. These averaged transitions are marked by a double dagger in Table S3 in Supplementary Note 4. We identified combination differences for 9 upper states reached by $P6\leftarrow P2$ transitions from the $\nu_3$ $2F_1$ and $3F_1$ states, as shown in Fig. 6c. For each upper state, the independently measured final state term values agree within their $3\sigma$ uncertainties. All the measured $P6\leftarrow P2$ transitions have been unambiguously assigned based on the ExoMol predictions with the difference between the measured center wavenumbers and the predicted values plotted in Fig. 6d, and their mean and standard deviation stated in Table 3. Moreover, a $P6\leftarrow P2$ line list from the recently developed new *ab initio*-based effective Hamiltonian approach[50] shows better agreement with the measurements than ExoMol, as illustrated in Table 3, with the mean wavenumber offset reduced by a factor of 7 and the scatter by a factor of 2, consistent with what was observed in this range in our previous work[33,51,52]. However, this effective Hamiltonian approach currently does not provide predictions at higher polyads due to the exponential growth in the number of sub-levels and severe resonant coupling, where the perturbative model breaks down.

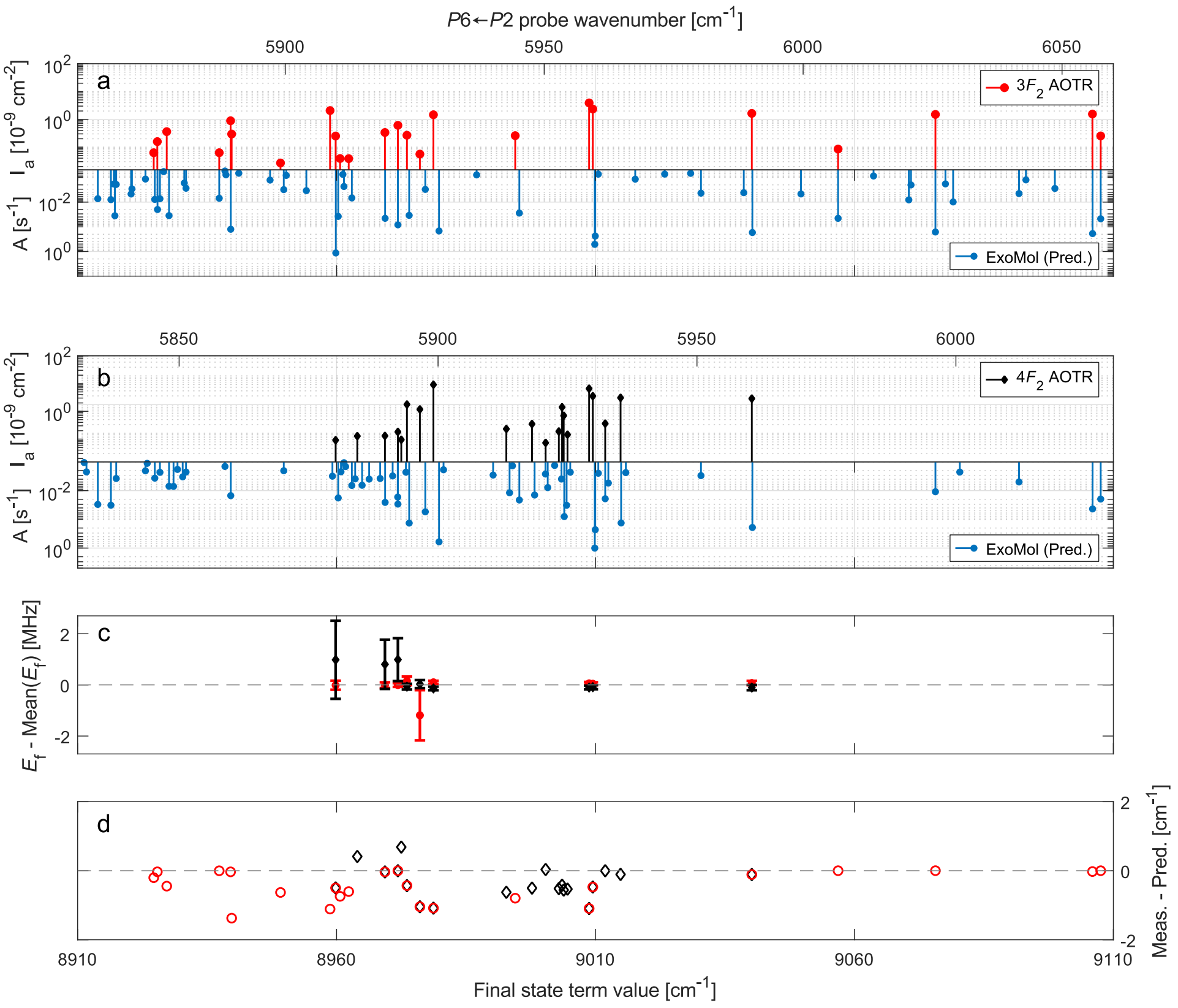

**Fig. 6. Measured *P*6←*P*2 transitions and comparison to ExoMol predictions.** Integrated absorption coefficients, $I_a$, of ladder-type *P*6←*P*2 transitions detected in **a** $3F_2$ AOTR measurement from the lower state $\nu_3$ $2F_1$ and **b** $4F_2$ AOTR from the lower state $\nu_3$ $3F_1$ are plotted in the upper panels with the corresponding Einstein A coefficients of the ExoMol predictions shown in the bottom panels, inverted for clarity. **c** The difference between the individual final state term values, $E_f$, in the *P*6 polyad and their weighted mean for the transitions that form combination differences. **d** Differences between the measured center wavenumbers and the ExoMol predictions of the *P*6←*P*2 transitions.

All V-type *P*4←*P*0 transitions in the three AOTR measurements originate from the same rovibrational sublevel of the ground state, and therefore some transitions were observed in two or three AOTR spectra in the overlapping probe ranges. Of the 15 distinct transitions, 14 were observed in the $4F_2$ AOTR spectrum, plotted in black in Figure 7a, and one was observed only in the $3F_2$ AOTR spectrum, plotted in red. The integrated absorption coefficient of the latter was scaled for plotting using the ratio of the integrated absorption coefficient of the strongest transition near 6004.6 $cm^{-1}$ in the $4F_2$ AOTR and $3F_2$ AOTR spectra. All V-type transitions observed in the $2F_2$ AOTR measurement were also observed in the $3F_2$ AOTR spectrum and are thus not plotted in Fig. 7a. In Table S4 in Supplementary Note 4, we list the mean probe wavenumbers weighted by the inverse squared uncertainties when a transition was observed in more than one measurement. The final state term values in the $P4$ polyad are calculated as the sum of the ground state term value and the measured wavenumbers of the V-type $P4 \leftarrow P0$ probe transitions. All measured V-type $P4 \leftarrow P0$ transitions except the one plotted with dashed line in Fig. 7a can be unambiguously assigned using four different line lists (see Table S4 in Supplementary Note 4): the ExoMol[5] (compared in Fig. 7b), the effective Hamiltonian[50], HITRAN2024[53], and WKLMC[54]. In all four cases, the assignments are straightforward to the closest predicted strong line (note that the WKLMC line list does not include all measured $P4 \leftarrow P0$ transitions, see Table S4 in Supplementary Note 4).

The line at 5949.61942(1) $cm^{-1}$, plotted with dashed line in the $4F_2$ AOTR measurement, is identified as a Λ-type transition[49] from the pumped $2\nu_3$ $4F_2$ state in the $P4$ polyad to the $5F_1$ ground state sublevel, based on the difference of the predicted $2\nu_3$ term values from ExoMol[5] and ground state value from de Oliveira *et al.*[49].

According to the mean values and standard deviations of the center wavenumber differences shown in Table 3, the prediction accuracy in the *P*4 polyad exceeds that in the *P*6 polyad by more than three orders of magnitude, as verified in our previous work[33,34,52]. The prediction accuracy further deteriorates in the *P*8 polyad, confirming that the accuracy of the theoretical model decreases significantly with increasing polyad number. The agreement in line-center wavenumbers is primarily limited by the accuracy of the theoretical model, as the differences exceed the measurement uncertainties by up to six orders of magnitude in the *P*6 and *P*8 polyads and by more than one order of magnitude in the *P*4 polyad. This trend highlights the utility of the AOTR experimental platform for providing high-accuracy spectral data about highly excited states.

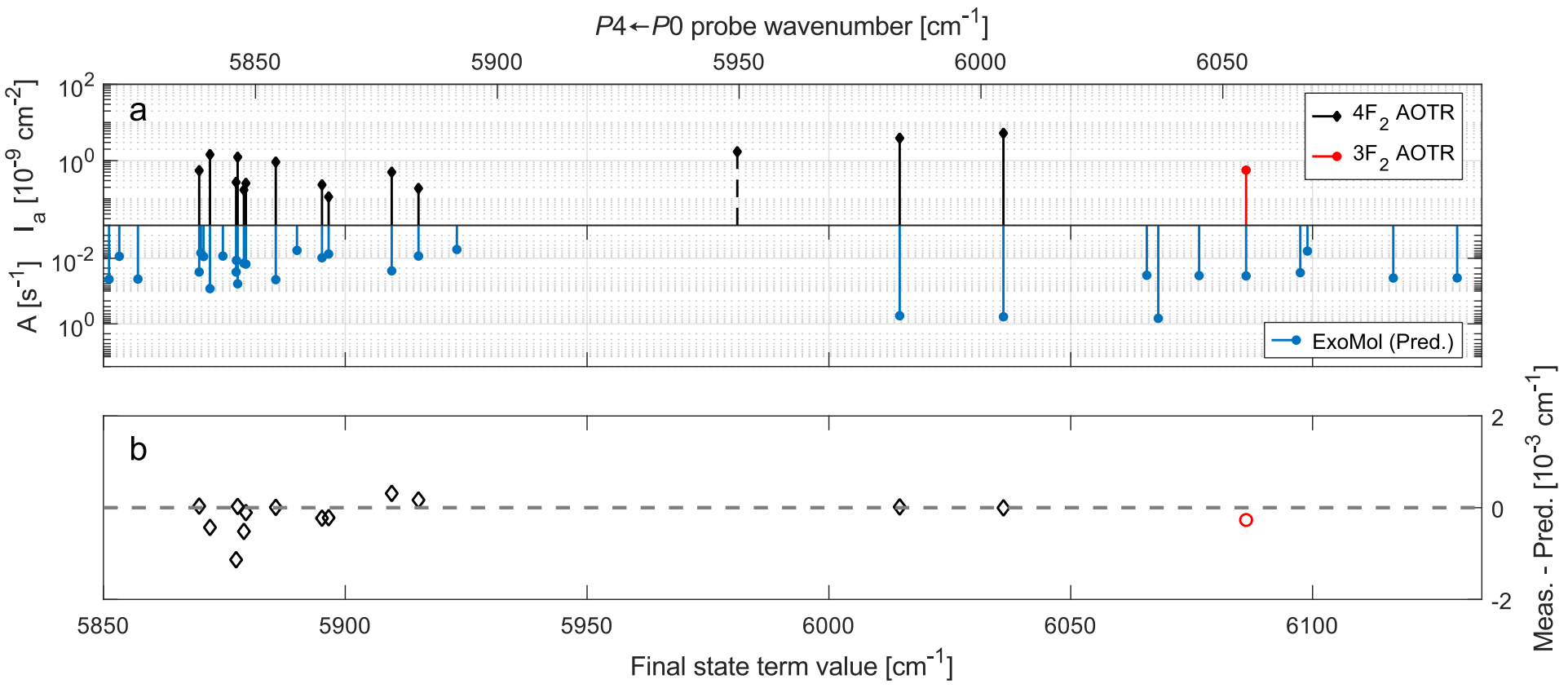

**Fig. 7. Measured V-type *P*4←*P*0 transitions and comparison to ExoMol predictions. a** V-type *P*4←*P*0 transitions observed in $4F_2$ AOTR (black) and $3F_2$ AOTR (red) measurements (see text for details). The transition shown by the dashed line is a Λ-type transition from the pumped $2\nu_3\,4F_2$ sublevel to the $5F_1$ ground state sublevel. **b** Differences between the measured center wavenumbers and the ExoMol predictions of the V-type *P*4←*P*0 transitions.

**Table 3.** Mean values and standard deviations of the discrepancies in center wavenumbers between the measurements and the predictions in the different spectral ranges.

| Transition range | Reference source | Mean offset [$cm^{-1}$] | Standard deviation [$cm^{-1}$] |
|---|---|---|---|
| $P8 \leftarrow P4$ | ExoMol[5] | -2.59 | 0.57 |
| $P6 \leftarrow P2$ | ExoMol[5] | -0.37 | 0.45 |
| | Hamiltonian[50] | 0.008 | 0.21 |
| V-type $P4 \leftarrow P0$ | ExoMol[5] | $-2 \times 10^{-4}$ | $3 \times 10^{-4}$ |
| | Hamiltonian[50] | $-2 \times 10^{-4}$ | $8 \times 10^{-4}$ |
| | HITRAN[53] | $-4 \times 10^{-4}$ | $3 \times 10^{-4}$ |
| | WKLMC[54] | $-1 \times 10^{-5}$ | $3 \times 10^{-4}$ |

# 3. Discussion and Outlook

In this work, we demonstrate AOTR spectroscopy using a two-step CW pump and cavity-enhanced comb probe that enables broadband detection and assignment of Doppler-free transitions between a wide range of vibrational states. The stepwise MIR pumping scheme opens a way to populate a specific velocity group of molecules to a broad range of intermediate excited states, while the broadband probing achieved by the NIR comb probe enables direct exploration over a broad spectral and energy range. Furthermore, combining single-pass MIR pumping with cavity-enhanced NIR probing decouples the pump and probe fields, enabling the pump wavelengths to be fixed for velocity-selective excitation while maintaining high probing sensitivity. These features overcome the key limitations of OODR spectroscopy using two CW cavity-enhanced NIR lasers[9], namely the limited spectral span and the coupling of pump and probe frequencies imposed by locking to the cavity.

An additional advantage of the triple-resonance scheme is its ability to extend the rotational angular momentum selection rule of a single transition ($\Delta J = 0, \pm1$) to an overall selection rule of $\Delta J = 0, \pm1, \pm2, \pm3$ between the ground and final states of the triple resonance scheme.

In the proof-of-concept demonstration of comb-based AOTR, we deliver the first set of high-accuracy and high-precision spectral data for highly excited states up to the $P8$ polyad of methane. At present, the low accuracy of the theoretical predictions in the $P8 \leftarrow P4$ range precludes line assignment based on closest wavenumber and intensity match. We employed the combination-difference approach to rotationally assign 27 out of 105 probed $P8 \leftarrow P4$ transitions, which then allowed full vibrational assignment using ExoMol predictions. 12 additional transitions were assigned based on the observed vibrational-band-dependent shifts. Further assignments will be possible by pumping other intermediate states in the $P4$ polyad and measuring polarization-dependent intensity ratios[31] using different relative pump/probe polarization configurations. The strength of the AOTR signals will be further increased to allow detection of weak $P8 \leftarrow P4$ transitions by optimizing the Rabi frequencies[27] of the two pump transitions to increase the population at the selected state in the $P4$ polyad.

The comb-stabilized double-seeded CW-OPO uniquely enables the generation of two broadly tunable, frequency-stabilized MIR idlers at arbitrary wavelengths within the gain bandwidth of the nonlinear crystal inside the CW-OPO, as demonstrated by the three different pairs of pump transitions, separated by up to 34 $cm^{-1}$, excited for *P*8←*P*4 combination-difference measurements. The current double-seeded CW-OPO has a tunability exceeding 80 $cm^{-1}$ by directly tuning the seed lasers, allowing pumping many transition pairs to populate different states within the *P*2 and *P*4 polyads (see Supplementary Note 1, Figure S1d), with further extension achievable by tuning the CW-OPO signal frequency. The power ratio between the two idlers can be readily changed by adjusting the power ratio of the seed lasers, while the two idlers inherently share the same spatial mode and perfectly overlap in free space, eliminating the need for external optical alignment; these two advantages significantly simplify the experimental setup. The comb-based frequency stabilization provides a flexible and reliable approach for locking multiple frequencies at arbitrary values, which is a superior alternative to conventional locking methods, such as locking the pump laser to a Lamb dip[11,31,32] or to weak ladder-type absorption features[10], both of which require frequency modulation and introduce sidebands in the probe spectrum. With these merits, the comb-stabilized double-seeded CW-OPO offers a powerful and versatile alternative for OODR spectroscopy compared to the previously reported experimental configurations using two QCLs[27], two independent CW-OPOs[55], or two independent difference frequency generation (DFG) processes[10], and is readily applicable to a broad range of needs requiring two or more frequency-stabilized MIR lasers. Ladder climbing to even higher states by seeding the CW-OPO pump amplifier with three or more frequencies should be possible.

AOTR spectroscopy, integrating a comb-stabilized double-seeded CW-OPO as the pump with a cavity-enhanced NIR comb as the probe, is a new room-temperature method that can be widely applied to investigate previously inaccessible highly excited rovibrational states of a wide range of molecules. Spectral data generated by this technique will fill critical gaps in databases and help refine theoretical models required for accurate interpretation of high-temperature molecular spectra in astrophysical and combustion environments.

## 4. Methods

### *4.1 Two-step pumping by double-seeded CW-OPO*

Two pump fields are provided by two idlers of a singly-resonant CW-OPO (TOPTICA Photonics, TOPO) whose pump amplifier is seeded by two narrow-linewidth 1.06 μm lasers (seed1: external-cavity diode laser, TOPTICA Photonics, CTL PRO, specified short-term linewidth < 10 kHz at 5 μs; seed2: ytterbium fiber laser, NKT Photonics, <20 kHz at 120 μs). The two seed lasers are combined using a polarization-maintaining fiber coupler, then amplified by an optical amplifier (YAR-10-1064-LP-SF, IPG Photonics), and sent to a singly-resonant CW-OPO, in which they generate a single signal wave at around 1.57 μm. The frequencies of the idlers ($\nu_{idler} = \nu_{seed} - \nu_{signal}$) are coarsely determined by simultaneously measuring the frequencies of seed lasers ($\nu_{seed}$) and CW-OPO signal ($\nu_{signal}$) using the NIR FTS with a frequency resolution of 250 MHz (the same as the one used for measuring probe spectra). The idler frequencies are coarsely tuned by adjusting the crystal position and etalon angle inside the CW-OPO to change the signal frequency and are finely tuned by tuning the two seed laser frequencies to match the pump frequencies listed in Table 1. The frequency stabilization of the idlers is achieved by phase locking to the MIR comb, produced by DFG from a single oscillator and thus free from carrier envelope offset frequency[56]. A small fraction of the idlers reflected by a beamsplitter is aligned to overlap with the MIR comb output by a pellicle beamsplitter, then dispersed by a reflection grating (G1), enabling the beat-note signals at two pump wavelengths to be detected by two photodiodes, PD1 and PD2, respectively. Setting the MIR comb repetition rate, $f_{rep}^{DFG}$, to 125.06 MHz, 125.08 MHz and 125.06 MHz for $2F_2$, $3F_2$ and $4F_2$ AOTR measurements, respectively, the beat note between pump1 and the nearest comb line, $f_{beat,1}^{DFG}$, is 20.10 MHz, 23.35 MHz and 20.34 MHz, respectively, while the corresponding beat note of pump2 with its nearest comb line, $f_{beat,2}^{DFG}$, is 58.17 MHz, 57.92 MHz and 59.43 MHz, respectively. Two phase detectors are employed to compare the detected beat-note signals with reference signals from RF1 (Menlo Systems, DDS120 Synthesizer) and RF2 (Analog Devices, AD9914), respectively, thereby generating error signals for phase-locking the beat notes at the target frequencies. All RF sources and counters used in the experiment are referenced to a GPS-disciplined rubidium clock (Stanford Research Systems, SRS FS725). The error signals are sent to proportional-integral (PI) controllers (New Focus, LB1005). The frequencies of the seed lasers are corrected by applying the feedback signals to the drive current of the external-cavity diode laser and the piezoelectric transducer (PZT) of the ytterbium fiber laser, respectively, consequently stabilizing the idlers at the target frequencies. During the measurement, both stabilized beat notes are monitored using a two-channel frequency counter (Rohde & Schwarz, HM8123, 1 s gate time, not shown in Fig. 1b). For the $2F_2$, $3F_2$ and $4F_2$ AOTR measurements, the Allan deviations of the counted $f_{beat,1}^{DFG}$ at 3.42 s integration time are 386 Hz, 5.42 kHz and 0.89 Hz, while those of $f_{beat,2}^{DFG}$ are 51 Hz, 480 Hz and 253 Hz, respectively. The individual power levels of pump1 and pump2 incident on the sample for the $2F_2$ and $4F_2$ AOTR measurements, given in Table 1, are determined from their measured power ratio. The two pump beams were spatially separated using a reflection grating (G3) placed behind the cavity, as shown in Fig. 2, taking advantage of their large wavenumber difference and hence large angular separation. For the $3F_2$ AOTR measurement, pump1 and pump2 have small wavenumber difference and therefore cannot be completely spatially separated by the grating. Thus, their individual power levels incident on the sample, listed in Table 1, are estimated by comparing the cavity transmission under vacuum and with 300 mTorr of $CH_4$ inside the cavity, where pump1 is fully depleted by the $\nu_3$ Q(2, $F_2$) transition and pump2 passes through.

We note that in the $3F_2$ AOTR measurement, pump2 was locked at the ExoMol-predicted line center of the $2\nu_3 \leftarrow \nu_3$ R(2, $F_1$) pump transition because we did this measurement before we measured the line center with sub-MHz uncertainty using the OODR method described in Supplementary Note 2. The measured line center of the $2\nu_3 \leftarrow \nu_3$ R(2, $F_1$) pump transition is 416(62) kHz higher than the ExoMol prediction (see Supplementary Note 2). We note that the pump offset from the transition does not shift the line center of the $P8 \leftarrow P4$ probe transitions since the cavity-enhanced probe beam is both co- and counter-propagating with respect to the pump and simultaneously interacts with two molecular velocity groups with opposite signs, which results in a broadening rather than shift of the probe transitions.[32]

### *4.2 Probe frequency stabilization and locking to the optical cavity*

The comb probe is an amplified Er:fiber frequency comb (Menlo Systems, FC1500-250-WG) with $f_{rep}$ = 250 MHz and centered at 1.55 μm. To cover the probe range around 6000 cm$^{-1}$, the comb is shifted to 1.68 μm by a polarization-maintaining microstructured silica fiber[57]. The comb modes are coupled to the 60-cm optical cavity with a finesse of 1220 at 5960 cm$^{-1}$, which extends the absorption length by a factor of 777. To lock the comb to the cavity via the PDH locking technique, the comb is phase-modulated at 20 MHz by an electro-optic modulator (EOM) inserted between the oscillator and the amplifier (not shown in Fig. 2). The cavity-reflected comb light is collected using an optical circulator, collimated onto a reflection grating (G2) in free space and dispersed to image the part at the selected wavelength on a photodiode (PD3). The PDH error signal is obtained by demodulating the output of PD3 at 20 MHz and sent to two PI controllers (New Focus, LB1005). The feedback signals are separately applied to the intracavity EOM and PZT inside the comb oscillator to lock the modes to the cavity at the locking point selected by the reflection grating. The comb-cavity coupling efficiency is optimized by adjusting the carrier envelope offset frequency of the comb, $f_{ceo}$, to compensate for the offset between the comb modes and the cavity modes. The $f_{ceo}$ is measured by an $f$-$2f$ interferometer and downshifted to be always locked at 20 MHz (RF3, Anritsu, MG3692A) by sending the feedback signal to the pump diode current of the comb oscillator. The $f_{rep}$ is referenced to a tunable RF source (RF4, Aim-TTi, TGF4162) by sending the feedback signal to the PZT attached to the rear sample cavity mirror. The $f_{rep}$ scan for spectral interleaving is achieved by tuning the frequency of RF4. The values of $f_{ceo}$ and $f_{rep}$ are monitored using a two-channel counter (Menlo Systems, FXM50) with a gate time of 1 s, which is not shown in Fig. 2. More details about the comb-cavity locking can be found in our previous work[34].

### *4.3 Spectral acquisition*

The cavity-transmitted comb probe is led to the FTS via a polarization-maintaining fiber. The optical path difference of the FTS is calibrated using a stabilized 633-nm HeNe laser (Sios, SL/02/1) with a fractional frequency stability of $5\times10^{-9}$ over 1 h. The comb interferograms are recorded simultaneously with the HeNe interferograms using a data acquisition card (National Instruments, PCI5922) with a sampling rate of 5 MS/s and 20-bit resolution. The sub-nominal resolution scheme[45] relies on resampling the spectral signal at comb mode positions to minimize the instrumental line shape in the spectra, which is achieved in post-processing by adjusting the HeNe laser wavelength, $\lambda_{ref}$. For spectral interleaving, the comb $f_{rep}$ is stepped with a step size of 2.75 Hz by tuning the frequency of RF4, resulting in about 2 MHz sample point spacing in the AOTR spectrum. To fill the frequency gaps between adjacent comb modes, 130 $f_{rep}$ steps are needed in a full scan, which takes about 16 min. For each step, the nominal resolution of the FTS is matched in post-processing to the $f_{rep}$, and the frequency scale is shifted by the $f_{ceo}$, which yields spectra with a comb-calibrated frequency axis[45]. At each $f_{rep}$ step, we record two interferograms with the pump fields unblocked (signal spectrum) and blocked (background spectrum) by an optical shutter, respectively. The AOTR spectrum is normalized by the background spectrum during post-processing. To improve the signal-to-noise ratio of the spectrum, we averaged 7, 14 and 10 series of $f_{rep}$ scans during the $2F_2$, $3F_2$ and $4F_2$ AOTR measurements, respectively, and 6 and 5 scans in the measurement with pump2 turned off and pump1 on for exciting $\nu_3$ Q(2, $F_2$) and $\nu_3$ R(2, $F_2$) pump transitions, respectively. In all measurements, $f_{rep}$ was stepped in alternating directions.

### *4.4 Frequency uncertainty*

The uncertainties in the probe transition center frequencies originate from three sources: the fit uncertainty, the sub-nominal resolution procedure and the pressure shift, and the total uncertainties are the quadrature sum of them. The fit uncertainties of the $P8 \leftarrow P4$ transitions varied between 45 kHz and 1.1 MHz, for the $P6 \leftarrow P2$ transitions between 29 kHz and 1.5 MHz, and for the $P4 \leftarrow P0$ transitions between 195 kHz and 1.2 MHz, depending on the signal-to-noise ratio of each transition. To evaluate the uncertainty contribution of $\lambda_{ref}$ optimization generated in the sub-nominal resolution procedure, we fitted four strong $P6 \leftarrow P2$ transitions in each probe spectrum from three AOTR measurements analyzed using different $\lambda_{ref}$ values around the optimum. The optimum $\lambda_{ref}$ is the one that corresponds to the smallest instrumental line shape function. We then made a linear fit to the transition line center frequencies obtained at different $\lambda_{ref}$ values. Based on the slope of this dependence and the uncertainty in the optimum $\lambda_{ref}$, we estimated center frequency uncertainties of 79 kHz ($2.6 \times 10^{-6}$ cm$^{-1}$), 20 kHz ($6.8 \times 10^{-7}$ cm$^{-1}$) and 44 kHz ($1.46 \times 10^{-6}$ cm$^{-1}$) introduced by the sub-nominal resolution procedure for the $2F_2$, $3F_2$ and $4F_2$ AOTR measurements, respectively.

The influence of the pressure shift on the probe transitions is estimated based on literature values. Using the self-induced pressure shift coefficient of ($-0.017 \pm 0.003$) cm$^{-1}$/atm$^{-1}$ for methane transitions in the 6000 cm$^{-1}$ range reported by Lyulin *et al.*[58], we estimate the pressure shift at 60 mTorr to be -40 kHz and include it in the frequency uncertainty budget. As discussed in our previous work[34], this value is probably overestimated; it is conservatively included here for consistency[32–34]. In addition, we estimated the power shift of the probe transitions using the power shift coefficient of -13 ± 17 kHz/W reported by Okubo *et al.*[46] for a Doppler-free Lamb dip in the $P(7, E)$ line of the $\nu_3$ band with a beam radius of 0.71 mm, similar to the pump beam radius of 0.7 mm in this work. For the total pump power levels incident on the gas sample of 462 mW, 550 mW and 614 mW used in $2F_2$, $3F_2$ and $4F_2$ AOTR measurements, respectively, the power shifts of the probe transitions in these measurements are -6.0 ± 7.9 kHz, -7.2 ± 9.4 kHz, and -8.0 ± 10.4 kHz, which are negligible compared to the pressure shift and the uncertainty from the sub-nominal resolution procedure. Moreover, a possible AC-Stark shift does not shift the line center in OODR detection[59] and is expected not to affect the AOTR detection either. The recoil shift and the second-order Doppler shifts of ladder-type transitions[60] are about 4.5 kHz and about -300 Hz, respectively, and can also be neglected.

The uncertainties in the final state term values in the $P8$ polyad (Table S2 in Supplementary Note 4) are calculated as the quadrature sum of the uncertainties of the probed $P8 \leftarrow P4$ transition wavenumbers described above, the 0.85 kHz uncertainty of the $2F_2$ ground state term value from de Oliveira *et al.*[49], the uncertainties of the $\nu_3$ pump transitions listed in Table 1 from Okubo *et al.*[46] and Abe *et al.*[47], the uncertainties of the $2\nu_3 \leftarrow \nu_3$ pump transitions listed in Table 1, yielding values between 134 kHz and 844 kHz for the $2F_2$ AOTR measurement, between 108 kHz and 1.1 MHz for the $3F_2$ AOTR measurement, and between 82 kHz and 987 kHz for the $4F_2$ AOTR measurement. The uncertainties in the final state term values in the $P6$ polyad (Table S3 in Supplementary Note 4) are calculated as the quadrature sum of the uncertainties in the probed $P6 \leftarrow P2$ transition centers, the same ground state term value and the $\nu_3$ pump transitions listed in Table 1, yielding values between 54 kHz and 1.5 MHz. The uncertainties in the final state term values in the $P4$ polyad (Table S4 in Supplementary Note 4) are calculated as the quadrature sum of the uncertainties in the probed $P4 \leftarrow P0$ transition centers and the same ground state term value, yielding values between 204 kHz and 1.2 MHz.

**Data Availability**: Machine-readable versions of the line lists with all transitions measured in this work are available at https://comblinelists.umu.se/molecules/.

**Acknowledgments**: The authors thank David A. Long and Adam T. Heiniger for useful discussions about double-seeded CW-OPOs, Piotr Masłowski for lending the rubidium clock, and Michael Rey for sharing the line list from his *ab initio* Hamiltonian predictions.

**Author contributions**: K.K.L. and A.F. conceived the idea. Q.N., V.S.O. and A.H. implemented the experiment and performed the measurements. Q.N., A.H., V.S.O. and I.S. analyzed the data. Q.N. visualized the data. K.K.L. and I.S. contributed theoretical predictions and analysis tools. All authors discussed the results. A.F. supervised the project and provided resources. Q.N. wrote and revised the manuscript, with extensive editing by A.F. All authors reviewed and commented on the manuscript.

**Funding**: Swedish Research Council (2020-00238, 2026-02023), Knut and Alice Wallenberg Foundation (KAW 2020.0303), Kempe Foundation (JCSMK24-0146, JCSMK261-0091), Wenner Gren Foundation (GFOv2024-0010), U.S. National Science Foundation (CHE-2108458).

**Competing interests**: The authors declare no competing interests.

# Triple-Resonance Spectroscopy Using a Cavity-Enhanced Frequency Comb Probe: Supplementary Information

**Qinxue Nie[1], Vinicius Silva de Oliveira[1], Adrian Hjältén[1], Isak Silander[1], Kevin K. Lehmann[2], and Aleksandra Foltynowicz[1,*]**

[1] Department of Physics, Umeå University, 901 87 Umeå, Sweden

[2] Departments of Chemistry & Physics, University of Virginia, Charlottesville, VA 22904, USA

[*]Corresponding author: aleksandra.foltynowicz@umu.se

## Supplementary Note 1: Spectral coverage of the two MIR CW idlers generated by the double-seeded CW-OPO

The comb-stabilized double-seeded CW-OPO developed in this work is used for populating different intermediate states in the *P*2 and *P*4 polyads. The CW-OPO is pumped by two NIR narrow-linewidth lasers, one of which is an external-cavity diode laser (ECDL, seed1) with a wide tuning range (1010-1100 nm), and the other is a fiber laser (seed2) with a limited tuning range (maximum of 680 pm, centered at 1064 nm). To determine the maximum possible frequency difference between the two idlers generated by the CW-OPO for covering a wide range of pump transition pairs, we tune the ECDL wavelength while the fiber laser wavelength remains fixed, as shown in Fig. S1a. The wavenumbers of both idlers are measured using a home-built MIR FTS with a resolution of about 0.06 $cm^{-1}$, as shown in Fig. S1c. The CW-OPO depleted seeds at 1.06 μm and the signal at 1.57 μm (Fig. S1b) leaking from the CW-OPO cavity are combined in a fiber coupler and aligned to the NIR FTS mentioned in the main text to simultaneously measure their wavenumbers with a resolution of 0.0083 $cm^{-1}$. As shown in Fig. S1b, the CW-OPO signal is maintained at a fixed wavenumber of 6383.5 $cm^{-1}$ in each measurement, close to that used in the AOTR measurement. Idler1 is tuned from 3067.5 $cm^{-1}$ to 2979 $cm^{-1}$, corresponding to the ECDL wavenumber being tuned from 9451 $cm^{-1}$ to 9362.5 $cm^{-1}$, while idler2 is located at around 3018 $cm^{-1}$, corresponding to the fiber laser wavenumber fixed at 9401.45 $cm^{-1}$. Within the wide tuning range of about 88.5 $cm^{-1}$, the single-mode operation of the CW-OPO idlers is maintained, enabling pumping two transitions with a large wavenumber difference. We note that the monitored depleted seeds exhibit irregular intensity variations during tuning; the underlying causes remain unclear and will be investigated quantitatively in future work.

Figure S1d shows the wavenumbers of possible transition pairs for stepwise pumping over a wide frequency range retrieved from the ExoMol database[1]. The predicted pump transition pairs are filtered using the conditions: $|\nu_{12} - \nu_{23}| < 60$ $cm^{-1}$, $A_{12} > 21$ $s^{-1}$, $A_{23} > 0.1$ $s^{-1}$ and $J_{max} = 10$, where $A_{12}$, $A_{23}$ are the Einstein A coefficients of the first pump transition and the second pump transition, respectively, $\nu_{12}$, $\nu_{23}$ are the optical frequencies, respectively, and $J_{max}$ is the maximum rotational quantum number. Note that it is possible to change the signal frequency to achieve a wider coverage of pump transition pairs. The three pump pairs used in this work for combination-difference measurements are marked in blue in Fig. S1d.

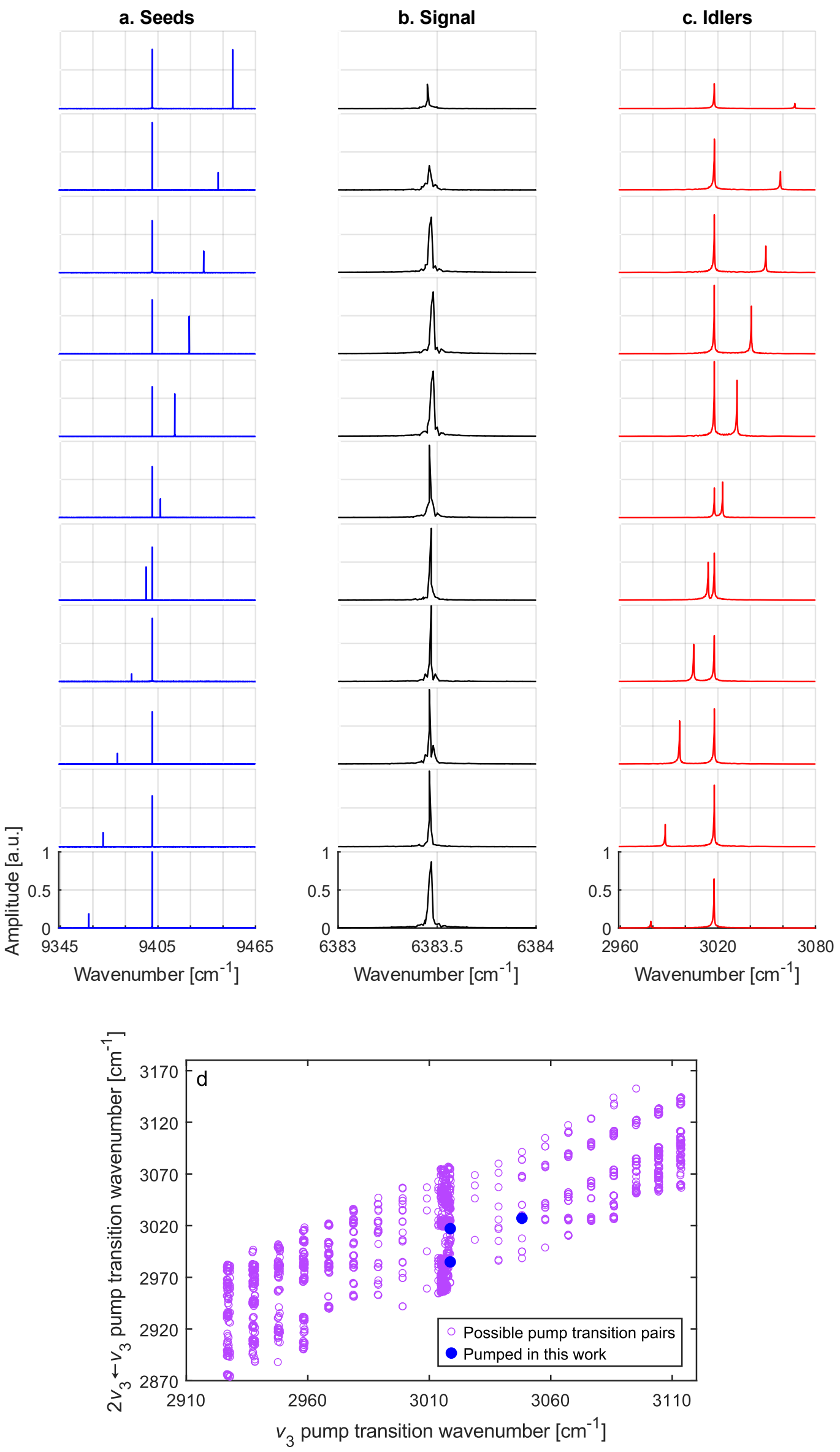

**Figure S1. Tunability of idlers generated by the double-seeded CW-OPO for covering a wide range of pump transition pairs. a** Wavenumbers of the seed lasers. **b** Wavenumbers of the CW-OPO signal, fixed to the value used in the $3F_2$ AOTR measurement. **c** Wavenumbers of the CW-OPO idlers. **d** Predicted pump transition pairs. The three blue points indicate the transition pairs pumped in this work.

## Supplementary Note 2: Line-center measurement of $2\nu_3 \leftarrow \nu_3$ pump transitions

We used CW MIR-MIR optical-optical double-resonance spectroscopy to determine the center frequencies of the $2\nu_3 \leftarrow \nu_3$ pump transitions with an accuracy better than that given in ExoMol[1]. The ExoMol predictions[1] of the $2\nu_3 \leftarrow \nu_3$ pump transitions used in the three AOTR measurements are listed in Table S1. For each line-center measurement, pump1 was locked to the center frequency of the corresponding $\nu_3$ pump transition listed in Table 1 of the main text, by referencing to the MIR comb. The cavity was filled with pure $CH_4$ at a pressure of 60 mTorr.

We used frequency modulation spectroscopy (FMS) to measure the line center of the $2\nu_3 \leftarrow \nu_3$ R(3, $F_1$) pump transition with sub-MHz accuracy. Pump2 (idler2) was frequency-modulated at 5.2 MHz using an electro-optic modulator between seed2 and the amplifier, while its center frequency was scanned across the $2\nu_3 \leftarrow \nu_3$ R(3, $F_1$) line by scanning seed2 frequency using the PZT. Figure S2a shows the FMS signal obtained by demodulating the signal of pump2 detected by the photodiode (PD4) at the cavity transmission. We locked the optical frequency of pump2 at the line center of the $2\nu_3 \leftarrow \nu_3$ R(3, $F_1$) transition, as indicated by the locking point in Fig. S2a, using the first harmonic signal as the error signal fed back to the PZT of seed2. The beat note between the locked pump2 and the MIR comb was recorded by a frequency counter for about 10 minutes, as shown in Fig. S2b. The mean value of the recorded beat note frequency was used to correct the ExoMol line center of the $2\nu_3 \leftarrow \nu_3$ R(3, $F_1$) transition and the standard deviation was used as the uncertainty. The slight drift in the beat note frequency was caused by the drift of the error signal caused by residual amplitude modulation (RAM) in the CW-OPO output. The corrected line center is provided in Table S1, which is about 23 MHz lower than the ExoMol prediction with a much lower uncertainty of about 34 kHz.

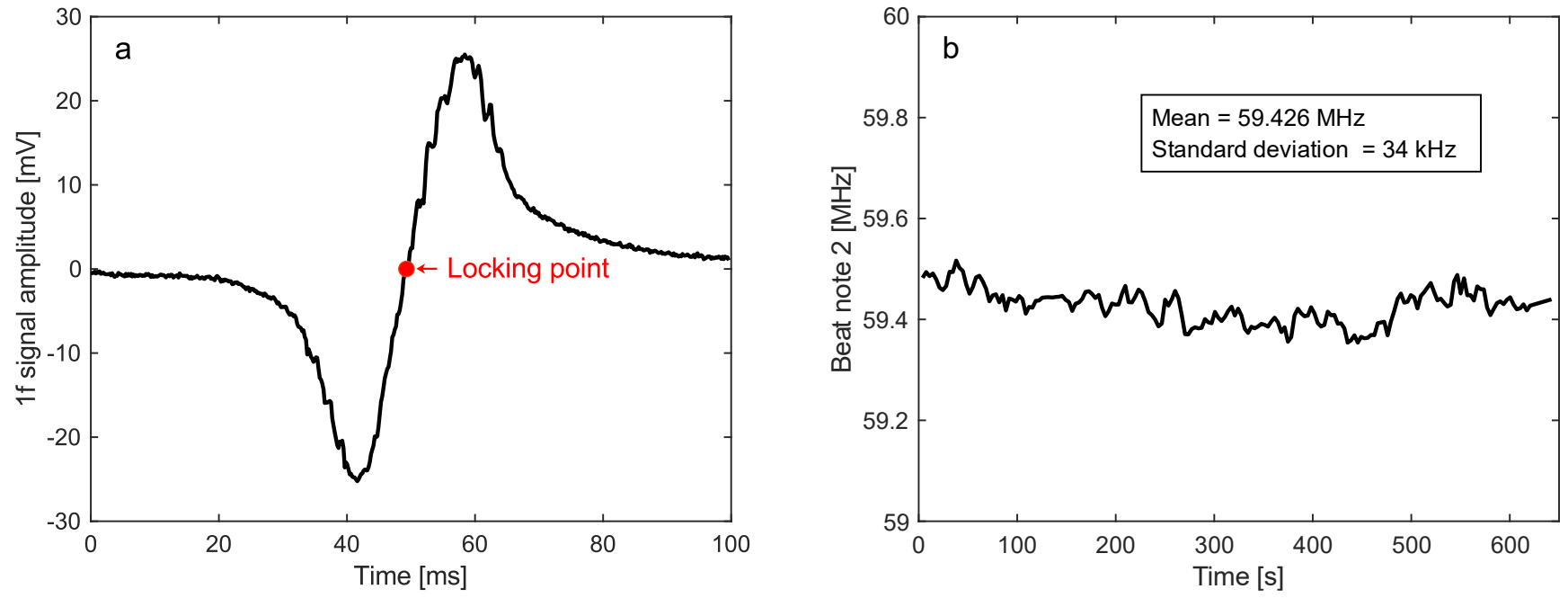

**Figure S2. Measurement of the line center of the $2\nu_3 \leftarrow \nu_3$ R(3, $F_1$) pump transition by frequency modulation spectroscopy**. **a** The FMS signal obtained by scanning the center frequency of pump2. **b** The continuously counted beat note frequency between pump2 with its nearest MIR comb line.

To determine the line centers of the 2ν₃←ν₃ Q(2, $F_1$) and R(2, $F_1$) transitions, we used wavelength modulation spectroscopy (WMS) instead of FMS, because they are much weaker than the $2\nu_3 \leftarrow \nu_3$ R(3, $F_1$) transition and their FMS signals are too strongly affected by the RAM of the CW-OPO output. We locked pump2 close to the ExoMol predicted line centers of these two $2\nu_3 \leftarrow \nu_3$ transitions by referencing to the MIR comb. The reference frequency generated by RF2, shown in Fig. 2 in the main text, used for locking the beat note frequency between pump2 and the MIR comb, was modulated at a frequency of 100 Hz and an amplitude of 1.5 MHz and its center frequency was stepwise scanned across the $2\nu_3 \leftarrow \nu_3$ pump transitions with a step size of 400 kHz and 27 steps. The phase locking between pump2 and the corresponding MIR comb line was maintained during the scanning. The WMS second-harmonic (2f) signals were obtained by demodulating the cavity transmitted pump2 detected by a photodetector at 200 Hz. The WMS signals of the $2\nu_3 \leftarrow \nu_3$ Q(2, $F_1$) and R(2, $F_1$) transitions, averaged 18 and 39 times, respectively, together with their fits are shown in Fig. S3a and b. The corrected line centers are provided in Table S1. The measured line center of the $2\nu_3 \leftarrow \nu_3$ Q(2, $F_1$) transition is 0.8 MHz higher than the ExoMol prediction with a much lower uncertainty of about 55 kHz. The measured line center of the $2\nu_3 \leftarrow \nu_3$ R(2, $F_1$) transition is 0.4 MHz higher than the ExoMol prediction with a much lower uncertainty of 62 kHz. All measured line center and uncertainties of the $2\nu_3 \leftarrow \nu_3$ pump transitions were used to calculate the final state term values measured in the three AOTR measurements.

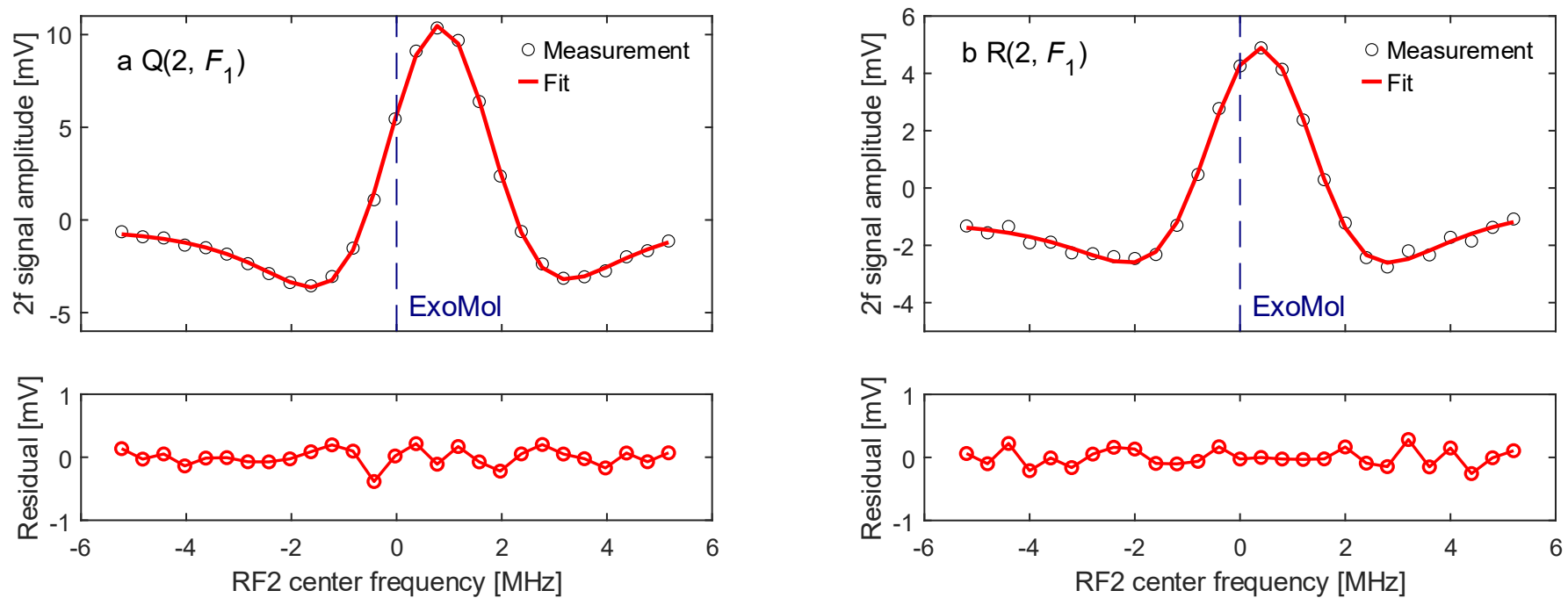

**Figure S3. WMS signals of the $2\nu_3 \leftarrow \nu_3$ pump transitions. a** Q(2, $F_1$) and **b** R(2, $F_1$) with fits (solid lines) and residuals (lower panels).

**Table S1. Comparison between the measured line centers of the $2\nu_3 \leftarrow \nu_3$ pump transitions and their ExoMol predictions.**

| $2\nu_3 \leftarrow \nu_3$ pump transition | ExoMol prediction [$cm^{-1}$] | Measurement [$cm^{-1}$] | Used in |
|---|---|---|---|
| Q(2, $F_1$)<br>(4 2 $F_2$ 80)←(2 2 $F_1$ 8) | 2984.878431(27)[1] | 2984.878457(2) | 2$F_2$ AOTR |
| R(2, $F_1$)<br>(4 3 $F_2$ 106)←(2 2 $F_1$ 8) | 3017.111829(27)[1] | 3017.111842(2) | 3$F_2$ AOTR |
| R(3, $F_1$)<br>(4 4 $F_2$ 140)←(2 3 $F_1$ 13) | 3027.163426(1000)[1] | 3027.162657(1) | 4$F_2$ AOTR |

**Supplementary Note 3: Line fitting procedures**

To retrieve the parameters of all Doppler-free ladder-type transitions, i.e., $P8 \leftarrow P4$ and $P6 \leftarrow P2$ transitions, we fit the cavity transmission function[2] to the averaged interleaved normalized spectra. In the cavity transmission function, as shown in Fig. S4a and Fig. S4b, the line shape was modeled as a sum of two profiles: a narrow and strong Lorentzian function for the Doppler-free probe transition, and a wider, weaker Gaussian function for the Doppler-broadened background absorption originating from thermal redistribution of the population of the upper pump level by elastic velocity-changing collisions[3]. The Doppler-broadened transitions from the ground state, simulated using the line parameters from HITRAN2024[4], were also included in the cavity transmission model. This model was divided by the transmission model that only contains the Doppler-broadened lines to account for the normalized step. This step was used to include the interference caused by the intracavity dispersion and gas absorption which modifies the effective cavity enhancement of the probe absorption length. In the fit, the center frequency of the transition, common for the Lorentzian and Gaussian components, the Lorentzian and Gaussian widths, the integrated absorption of each component, and the comb-cavity phase offset were set as free parameters. The remaining slight baseline distortions were fitted by a polynomial function with an order below 3. The fitting range for the $P6 \leftarrow P2$ ladder-type transitions was set to ±450 MHz, whereas that for the $P8 \leftarrow P4$ ladder-type transitions was set to ±200 MHz because their linewidths are around half those of the $P6 \leftarrow P2$ transitions. For some lines with strong $2\nu_3$ Doppler-broadened background, the fitting range was reduced to avoid baseline problems on the wings of the lines, because the Doppler-broadened features were not completely canceled by the normalization and not modeled correctly by the baseline. For the $P6 \leftarrow P2$ lines where the Gaussian component had a signal-to-noise ratio below 5, the Gaussian-to-Lorentzian integrated intensity ratio was fixed to 0.7, 0.6 and 0.5 and the Gaussian half-width at half maximum (HWHM) was fixed to 119, 91 and 114 MHz for the $2F_2$, $3F_2$ and $4F_2$ AOTR measurements, respectively, corresponding to the mean values obtained from fits to $P6 \leftarrow P2$ transitions with a signal-to-noise ratio higher than 5. For the $P8 \leftarrow P4$ lines, only one strong line exhibited a Gaussian component with a signal-to-noise ratio greater than 5. The fitted Gaussian HWHM of it is 25 MHz, and the Gaussian-to-Lorentzian intensity ratio is 0.5, which were used as fixed values for fitting other $P8 \leftarrow P4$ lines. A first-order polynomial baseline was used for 70% of the $P8 \leftarrow P4$ lines and 74% of the $P6 \leftarrow P2$ lines. A third-order polynomial was used for the remaining lines that overlapped with Doppler-broadened transitions.

To retrieve the parameters of V-type $P4 \leftarrow P0$ transitions, we fit them in the interleaved baseline-corrected transmission spectra without normalization using the same model as of the ladder-type transitions but with an inverted sign, as shown in Fig. S4c, as the V-type signal points in the opposite direction. Since all V-type transitions have a relatively low signal-to-noise ratio, we fixed the Gaussian width and the ratio between the Lorentzian and Gaussian integrated absorption coefficients to the mean values obtained from fits to seven strong $P6 \leftarrow P2$ transitions. In addition, we also fixed the Lorentzian width to the mean value of the HWHMs of strong $P6 \leftarrow P2$ transitions mentioned above for the corresponding measurement, since their widths should be equal[5]. The fitting range for V-type transitions was set to ±200 MHz. A third-order polynomial baseline was used for all V-type transitions. Figure 4 in the main text shows the fit results zoomed in to a ±100 MHz range around the line center for clarity.

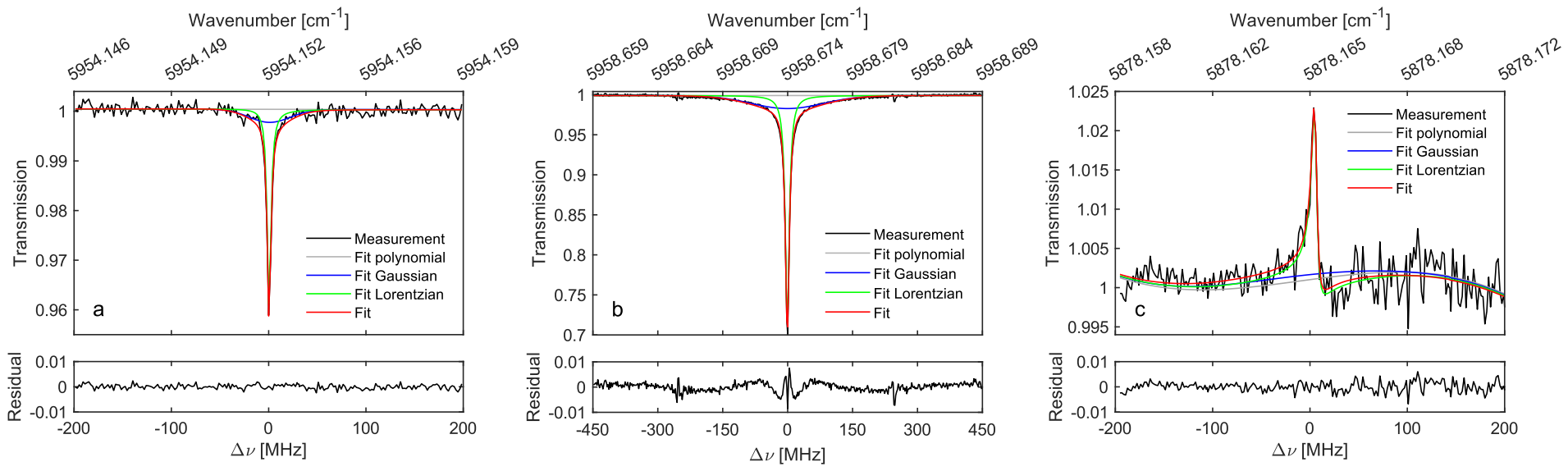

**Figure S4. Measured spectral signals of Doppler-free ladder-type and V-type transitions with fit models from 3$F_2$ AOTR measurement. a** A $P8 \leftarrow P4$ transition at 5954.152 cm$^{-1}$, **b** a $P6 \leftarrow P2$ transition at 5958.674 cm$^{-1}$ and **c** a V-type $P4 \leftarrow P0$ transition at 5878.165 cm$^{-1}$ with fits (red) generated using the fitting model comprised of a Gaussian component (blue) and a Lorentzian component (green). Fit residuals are shown in the lower panels.

**Supplementary Note 4: Line lists for all probe transitions**

Table S2 contains the fit results for all probed $P8 \leftarrow P4$ transitions with 39 of them assigned based on the combination-difference measurements and ExoMol predictions. The uncertainties of the probe transition wavenumbers are calculated as the quadrature sum of the fit uncertainties, the pressure shift and the uncertainties generated in the sub-nominal resolution procedure, as described in the main text Section 4.4. The uncertainties of final state term values are calculated as described in Section 4.4 in the main text. Note that for the $P8 \leftarrow P4$ transitions, we classify the assignments based on the available experimental evidence: assignments supported by both combination-difference measurements and vibrational-band-dependent shifts are considered more certain, whereas assignments based on only vibrational-band-dependent shifts are considered less certain, and are marked with a section sign (§) in Table S2. The upper state assignments ($P$ $J$ $C$ $N$) based on ExoMol predictions are shown in Column 6, where $P$ is the polyad number, $J$ the total angular momentum, $C$ the rovibrational symmetry, and $N$ the counting number. The vibrational assignments ($n_1$ $n_2$ $n_3$ $n_4$ $C_v$) from ExoMol are shown in Column 7, where $n_1$, $n_2$, $n_3$, $n_4$ are the vibrational modes, and $C_v$ the vibrational symmetry.

Table S3 contains the fit results for all probed $P6 \leftarrow P2$ transitions with comparison to predictions from ExoMol[1] and from the effective Hamiltonian approach[6]. The weighted mean probe center wavenumbers of the $P6 \leftarrow P2$ transitions observed in both $2F_2$ and $3F_2$ AOTR measurements are marked with a double dagger and used for calculating the corresponding final state term values. Among the observed transitions, four transitions marked with an asterisk were also measured in our previous work[3]. The center wavenumbers of these four transitions measured in this work lie within the uncertainty ranges of the previous results and exhibit improved precision, with uncertainties reduced to 10%–40% of those reported previously.

Table S4 contains the fit results for all probed V-type $P4 \leftarrow P0$ transitions. For transitions measured multiple times in different measurements, the weighted mean probe transition wavenumbers are reported and used for calculating the corresponding final state term values. They are compared to predictions from four available line lists: ExoMol[1], the effective Hamiltonian[6], HITRAN2024[4], and WKLMC[7]. Since the WKLMC line list does not cover the full measurement range, seven transitions do not have corresponding WKLMC assignments. The transition marked with an asterisk was also reported in our previous work[8]. The center wavenumber measured in this work lies within the uncertainty range of the previous result and exhibits improved precision, with the uncertainty reduced to 20% of that reported previously.

Table S5 lists all assigned final states in *P*8, *P*6 and *P*4 measured in this work, arranged in descending order of term value and compared with ExoMol predictions[1]. For the states reached by different pump-probe combinations, the weighted mean of all measured final state term values with the weighted uncertainties are listed.

**Table S2. List of all $P8 \leftarrow P4$ transitions detected in our work.** Column 1 identifies the corresponding measurement, pump transitions and the pumped $2\nu_3$ state term value in the $P4$ polyad. Column 2 lists the measured center wavenumbers, with the uncertainties given in parentheses. Column 3 shows the final state term values with uncertainties given in parentheses. Column 4 states the integrated absorption coefficients of the Lorentzian part of the probe transitions. Column 5 shows the HWHMs of the probe transitions. Column 6 gives the ExoMol upper-state assignments ($P\ J\ C\ N$), where $P$ is the polyad number, $J$ the total angular momentum, $C$ the rovibrational symmetry, and $N$ the counting number. Column 7 gives the ExoMol vibrational assignments ($n_1\ n_2\ n_3\ n_4\ C_v$), where $n_1$, $n_2$, $n_3$, $n_4$ are vibrational modes, and $C_v$ the symmetry. Column 8 gives the predicted Einstein A coefficients predicted by ExoMol. Column 9 gives the differences between the measured probe transition wavenumbers and ExoMol predictions. The transitions tentatively assigned based on the vibrational-band-dependent shifts in the absence of combination differences are marked with a section sign (§).

| 1 | 2 | 3 | 4 | 5 | 6 | 7 | 8 | 9 |
|---|---|---|---|---|---|---|---|---|
| Pump transitions and term value of pump state in $P4$ [$cm^{-1}$] | Probe transition wavenumber [$cm^{-1}$] | Final state term value [$cm^{-1}$] | Probe transition integrated absorption coefficient [$10^{-9}$ $cm^{-2}$] | Probe transition width [MHz] | Upper state assignment ($P\ J\ C\ N$) **ExoMol** | Vibrational assignment ($n_1\ n_2\ n_3\ n_4\ C_v$) **ExoMol** | Einstein A coefficient [$s^{-1}$] **ExoMol** | Meas.- **ExoMol** Pred. transition wavenumber [$cm^{-1}$] |
| Measurement:<br>$2F_2$ AOTR<br><br>Pump1:<br>$\nu_3$ Q(2, $F_2$):<br>3018.65020715(7)[9]<br>Pump2:<br>$2\nu_3 \leftarrow \nu_3$ Q(2, $F_1$)<br>2984.878457(2)<br><br>Term value of $2F_2$ in $P4$: 6034.971054(2) | 5867.64450(1) | 11902.61555(1) | 0.13(2) | 2.0(5) | | | | |
| | 5870.01221(3) | 11904.98326(3) | 0.11(2) | 3.3(8) | | | | |
| | 5871.675823(9) | 11906.646876(9) | 0.23(1) | 2.6(2) | | | | |
| | 5877.92844(2) | 11912.89950(2) | 0.047(8) | 2.3(6) | | | | |
| | 5879.520069(7)§ | 11914.491122(7) | 0.47(2) | 2.4(2) | 8 2 $F_1$ 1358 | 1 4 1 0 $F_2$ | 0.15939 | -3.6768 |
| | 5880.81492(2) | 11915.78597(2) | 0.11(2) | 2.2(5) | | | | |
| | 5881.51784(3) | 11916.48890(3) | 0.034(6) | 2.6(8) | | | | |
| | 5885.98051(2) | 11920.95157(2) | 0.062(8) | 2.6(5) | | | | |
| | 5886.961502(5) | 11921.932555(5) | 0.290(8) | 2.6(1) | | | | |
| | 5889.29428(2) | 11924.26533(2) | 0.036(6) | 2.5(6) | | | | |
| | 5892.24345(1) | 11927.21450(1) | 0.040(5) | 1.9(5) | | | | |
| | 5902.607133(8) | 11937.578187(8) | 0.111(6) | 2.7(2) | | | | |
| | 5903.84186(1) | 11938.81291(1) | 0.058(5) | 2.2(3) | | | | |
| | 5905.75257(2) | 11940.72363(2) | 0.029(5) | 2.2(6) | | | | |
| | 5909.378905(9) | 11944.349959(9) | 0.112(7) | 2.6(2) | | | | |
| | 5913.45375(1) | 11948.42480(1) | 0.076(6) | 2.3(3) | 8 3 $F_1$ 1956 | 1 4 1 0 $F_2$ | 0.034383 | -3.4585 |
| | 5913.470723(4) | 11948.441777(5) | 0.314(7) | 2.50(8) | | | | |
| | 5913.85756(1) | 11948.82862(1) | 0.068(8) | 2.6(4) | | | | |
| | 5915.365969(4) | 11950.337023(4) | 0.319(7) | 2.56(9) | 8 2 $F_1$ 1381 | 0 5 1 1 $F_2$ | 0.14847 | -2.5258 |
| | 5917.902129(5) | 11952.873182(6) | 0.228(9) | 2.3(1) | 8 2 $F_1$ 1383 | 0 5 1 1 $F_2$ | 0.14628 | -2.0121 |
| | 5924.824796(9) | 11959.795850(9) | 0.124(7) | 2.8(3) | 8 2 $F_1$ 1386 | 0 0 4 0 $F_2$ | 0.0073858 | -2.7918 |
| | 5926.13739(2) | 11961.10844(2) | 0.11(1) | 3.7(5) | 8 2 $F_1$ 1388 | 0 0 4 0 $F_2$ | 0.011185 | -2.583 |
| | 5929.628843(5) | 11964.599897(5) | 0.39(1) | 2.4(1) | 8 2 $F_1$ 1391 | 0 2 3 0 $F_1$ | 0.10496 | -2.4094 |
| | 5930.37042(2) | 11965.34147(2) | 0.07(1) | 2.6(6) | | | | |
| | 5936.685945(6)§ | 11971.656999(6) | 0.134(7) | 2.2(2) | 8 1 $F_1$ 877 | 0 0 4 0 $E$ | 0.33993 | -1.9569 |
| | 5947.73524(2) | 11982.70629(2) | 0.058(9) | 2.7(7) | 8 3 $F_1$ 1985 | 0 5 1 1 $F_2$ | 0.014541 | -2.5093 |
| | 5957.05132(2) | 11992.02237(2) | 0.14(1) | 3.3(5) | 8 2 $F_1$ 1399 | 0 0 4 0 $E$ | 0.042524 | -1.965 |
| | 5957.88074(2) | 11992.85180(2) | 0.07(2) | 1.7(4) | | | | |
| | 5961.62116(1) | 11996.59222(1) | 0.054(9) | 1.9(6) | | | | |
| | 5962.69713(2) | 11997.66818(2) | 0.08(1) | 2.4(5) | | | | |
| | 5986.385673(9) | 12021.356726(9) | 0.21(1) | 2.5(3) | 8 3 $F_1$ 2011 | 0 0 4 0 $E$ | 0.098393 | -1.99 |
| Measurement:<br>3$F_2$ AOTR<br><br>Pump1:<br>$\nu_3$ Q(2, $F_2$)<br>3018.65020715(7)[9]<br><br>Pump2:<br>2$\nu_3$←$\nu_3$ R(2, $F_1$)<br>3017.111842(2)<br><br>Term value of 3$F_2$ in<br>*P*4: 6067.204439(2) | 5877.38714(4) | 11944.59158(4) | 0.027(6) | 4(1) | | | | |
| | 5881.22040(2) | 11948.42484(2) | 0.09(2) | 2.5(5) | 8 3 $F_1$ 1956 | 1 4 1 0 $F_2$ | 0.079881 | -3.4584 |
| | 5883.13261(1) | 11950.33705(1) | 0.040(6) | 1.5(3) | 8 2 $F_1$ 1381 | 0 5 1 1 $F_2$ | 0.066386 | -2.5258 |
| | 5884.51637(2) | 11951.72081(2) | 0.030(4) | 2.8(6) | | | | |
| | 5885.668748(6) | 11952.873186(6) | 0.071(6) | 1.9(3) | 8 2 $F_1$ 1383 | 0 5 1 1 $F_2$ | 0.066754 | -2.0121 |
| | 5892.59140(1) | 11959.79584(1) | 0.033(4) | 2.3(4) | 8 2 $F_1$ 1386 | 0 0 4 0 $F_2$ | 0.021886 | -2.7918 |
| | 5893.904014(9) | 11961.108452(9) | 0.095(7) | 2.3(3) | 8 2 $F_1$ 1388 | 0 0 4 0 $F_2$ | 0.090837 | -2.583 |
| | 5897.39546(1) | 11964.59990(1) | 0.044(4) | 2.2(3) | 8 2 $F_1$ 1391 | 0 2 3 0 $F_1$ | 0.021224 | -2.4094 |
| | 5911.798508(5) | 11979.002946(5) | 0.098(3) | 2.3(1) | | | | |
| | 5915.26910(2) | 11982.47354(2) | 0.023(5) | 2.5(6) | | | | |
| | 5915.50182(1) | 11982.70626(1) | 0.051(5) | 2.9(4) | 8 3 $F_1$ 1985 | 0 5 1 1 $F_2$ | 0.038364 | -2.5093 |
| | 5915.90260(2) | 11983.10704(2) | 0.020(4) | 2.9(9) | 8 3 $F_1$ 1986 | 0 0 4 0 $F_2$ | 0.012713 | -3.0025 |
| | 5916.47312(1) | 11983.67756(1) | 0.031(3) | 2.0(2) | | | | |
| | 5918.325584(6) | 11985.530022(6) | 0.083(4) | 2.5(2) | | | | |
| | 5918.68678(1) | 11985.89122(1) | 0.038(4) | 2.7(5) | | | | |
| | 5919.72306(1) | 11986.92750(1) | 0.035(4) | 2.7(5) | | | | |
| | 5922.346793(8) | 11989.551231(8) | 0.045(3) | 1.9(2) | | | | |
| | 5923.77645(1) | 11990.98089(1) | 0.047(4) | 2.3(3) | | | | |
| | 5924.097462(5) | 11991.301900(6) | 0.083(3) | 2.5(2) | | | | |
| | 5924.51014(1)§ | 11991.71457(1) | 0.029(3) | 1.9(3) | 8 2 $F_1$ 1399 | 0 0 4 0 $E$ | 0.015362 | -3.4207 |
| | 5924.81790(3) | 11992.02234(3) | 0.020(4) | 3.0(8) | 8 4 $F_1$ 2458 | 1 4 1 0 $F_2$ | 0.058819 | -1.9651 |
| | 5925.28696(1)§ | 11992.49140(1) | 0.031(3) | 2.2(3) | 8 4 $F_1$ 2459 | 1 4 1 0 $F_2$ | 0.035685 | -3.2897 |
| | 5928.24147(2)§ | 11995.44591(2) | 0.007(3) | 2(1) | 8 3 $F_1$ 1998 | 0 0 4 0 $F_2$ | 0.0065005 | -3.0619 |
| | 5930.07830(3) | 11997.28273(3) | 0.019(4) | 2.4(7) | | | | |
| | 5930.50141(2) | 11997.70584(2) | 0.027(5) | 2.0(5) | | | | |
| | 5931.74028(2) | 11998.94472(2) | 0.021(4) | 2.8(7) | | | | |
| | 5947.64985(2) | 12014.85428(2) | 0.020(5) | 2.1(7) | | | | |
| | 5952.256483(6) | 12019.460921(6) | 0.077(4) | 2.2(2) | | | | |
| | 5952.875679(5)§ | 12020.080117(6) | 0.145(5) | 2.8(2) | 8 4 $F_1$ 2489 | 0 0 4 0 $F_2$ | 0.13519 | -3.095 |
| | 5954.152300(3) | 12021.356739(4) | 0.212(4) | 2.26(7) | 8 3 $F_1$ 2011 | 0 0 4 0 $E$ | 0.22453 | -1.9899 |
| | 5954.422298(9) | 12021.626736(9) | 0.031(4) | 1.6(5) | | | | |
| | 5958.17659(1) | 12025.38103(1) | 0.055(5) | 2.4(3) | | | | |
| | 5959.120028(4) | 12026.324466(5) | 0.105(4) | 2.3(1) | | | | |
| | 5959.781712(3) | 12026.986151(4) | 0.154(4) | 2.25(9) | 8 4 $F_1$ 2497 | 0 5 1 1 $F_2$ | 0.10658 | -3.004 |
| | 5962.19512(1) | 12029.39956(1) | 0.035(4) | 2.0(4) | 8 4 $F_1$ 2499 | 0 5 1 1 $F_2$ | 0.010711 | -1.7869 |
| | 5968.39566(2) | 12035.60010(2) | 0.076(8) | 3.1(5) | | | | |
| | 5969.13767(1) | 12036.34211(1) | 0.063(7) | 2.4(4) | | | | |
| | 5969.51427(2) | 12036.71871(2) | 0.033(5) | 1.8(4) | | | | |
| | 5982.82897(3) | 12050.03341(3) | 0.031(7) | 2.6(9) | | | | |
| | 5995.000150(7)§ | 12062.204589(8) | 0.23(2) | 2.2(3) | 8 4 $F_1$ 2532 | 0 0 4 0 $E$ | 0.11239 | -2.1152 |

| | | | | | | | | | |
|---|---|---|---|---|---|---|---|---|---|
| | 6048.98513(2)§ | 12116.18957(2) | 0.15(2) | 2.5(5) | 8 3 $F_1$ 2063 | 0 4 2 0 $A_1$ | 0.18193 | -1.3801 |
| Measurement:<br>4$F_2$ AOTR<br><br>Pump1:<br>$\nu_3$ R(2, $F_2$)<br>3048.15331811(8)[10]<br><br>Pump2:<br>$2\nu_3 \leftarrow \nu_3$ R(3, $F_1$)<br>3027.162657(1)<br><br>Term value of 4$F_2$ in<br>$P4$: 6106.758364(1) | 5834.91463(3) | 11941.67299(3) | 0.20(8) | 2.0(6) | | | | |
| | 5841.66645(1) | 11948.42482(1) | 0.33(5) | 2.9(4) | 8 3 $F_1$ 1956 | 1 4 1 0 $F_2$ | 0.12569 | -3.4577 |
| | 5843.698451(9) | 11950.456816(9) | 0.33(4) | 2.3(3) | | | | |
| | 5844.14307(1) | 11950.90144(1) | 0.20(2) | 2.8(4) | | | | |
| | 5852.38629(3) | 11959.14466(3) | 0.06(2) | 2(1) | | | | |
| | 5858.21651(2) | 11964.97488(2) | 0.053(8) | 2.1(5) | | | | |
| | 5860.48588(2) | 11967.24425(2) | 0.36(4) | 3.6(5) | | | | |
| | 5861.88269(1) | 11968.64106(1) | 0.076(6) | 2.9(4) | | | | |
| | 5861.92375(2) | 11968.68212(2) | 0.056(7) | 2.7(5) | | | | |
| | 5872.17565(3) | 11978.93402(3) | 0.043(8) | 2.9(8) | | | | |
| | 5873.369311(7) | 11980.127675(7) | 0.090(5) | 2.2(2) | | | | |
| | 5874.40866(2)§ | 11981.16703(2) | 0.05(1) | 2.3(6) | 8 3 $F_1$ 1984 | 0 0 4 0 $F_2$ | 0.056958 | -3.0461 |
| | 5875.947902(3) | 11982.706266(3) | 0.337(7) | 2.45(7) | 8 3 $F_1$ 1985 | 0 5 1 1 $F_2$ | 0.15955 | -2.5085 |
| | 5876.348644(2) | 11983.107008(3) | 0.73(2) | 2.49(6) | 8 3 $F_1$ 1986 | 0 0 4 0 $F_2$ | 0.17302 | -3.0017 |
| | 5876.67911(3) | 11983.43747(3) | 0.028(6) | 2.8(9) | | | | |
| | 5877.15635(2) | 11983.91471(2) | 0.044(6) | 2.5(5) | | | | |
| | 5877.30372(2) | 11984.06208(2) | 0.032(6) | 2.3(7) | | | | |
| | 5889.19801(1) | 11995.95638(1) | 0.067(7) | 2.7(4) | | | | |
| | 5889.47572(3) | 11996.23408(3) | 0.030(6) | 3(1) | | | | |
| | 5901.85546(2) | 12008.61383(2) | 0.071(9) | 3.8(7) | | | | |
| | 5904.46893(1) | 12011.22730(1) | 0.062(7) | 2.2(3) | | | | |
| | 5914.598372(4) | 12021.356736(4) | 0.290(7) | 2.41(9) | 8 3 $F_1$ 2011 | 0 0 4 0 $E$ | 0.094444 | -1.9892 |
| | 5920.22778(2) | 12026.98614(2) | 0.037(8) | 1.6(4) | 8 4 $F_1$ 2497 | 0 5 1 1 $F_2$ | 0.043654 | -3.0032 |
| | 5922.64120(1) | 12029.39956(1) | 0.082(6) | 2.3(2) | 8 4 $F_1$ 2499 | 0 5 1 1 $F_2$ | 0.016341 | -1.7861 |
| | 5926.024647(9) | 12032.783011(9) | 0.047(6) | 1.8(4) | | | | |
| | 5932.44735(1) | 12039.20572(1) | 0.074(7) | 2.7(4) | | | | |
| | 5933.044566(9)§ | 12039.802930(9) | 0.083(6) | 2.0(2) | 8 4 $F_1$ 2513 | 0 0 4 0 $F_2$ | 0.033202 | -3.0545 |
| | 5933.53597(1) | 12040.29433(1) | 0.054(7) | 1.6(4) | | | | |
| | 5936.29499(2) | 12043.05335(2) | 0.051(6) | 3.4(7) | | | | |
| | 5944.014827(9) | 12050.773191(9) | 0.092(6) | 2.5(2) | | | | |
| | 5954.24631(1)§ | 12061.00467(1) | 0.118(8) | 2.9(3) | 8 4 $F_1$ 2531 | 0 0 4 0 $E$ | 0.15157 | -2.1073 |
| | 5969.50702(1) | 12076.26539(1) | 0.14(1) | 2.6(3) | | | | |
| | 5970.66562(2)§ | 12077.42399(2) | 0.08(2) | 2.1(7) | 8 5 $F_1$ 3093 | 0 5 1 1 $F_2$ | 0.027941 | -2.8023 |

**Table S3. List of all $P6 \leftarrow P2$ transitions detected in our work.** Column 1 identifies the corresponding measurement, pump transitions and the pumped $\nu_3$ state term value in the $P2$ polyad. Column 2 lists the measured center wavenumbers, with the uncertainties given in parentheses. Column 3 shows the final state term values with uncertainties given in parentheses. Column 4 states the integrated absorption coefficients of the Lorentzian part of the probe transitions. Column 5 shows the HWHMs of the probe transitions. Column 6 gives the ExoMol upper-state assignments ($P\ J\ C\ N$), where $P$ is the polyad number, $J$ the total angular momentum, $C$ the rovibrational symmetry, and $N$ the counting number. Two transitions have different counting numbers in the ExoMol and the Hamiltonian line lists; the counting numbers from the Hamiltonian are given in parentheses next to the values from ExoMol. Column 7 gives the ExoMol vibrational assignments ($n_1\ n_2\ n_3\ n_4\ C_v$), where $n_1$, $n_2$, $n_3$, $n_4$ are vibrational modes, and $C_v$ the vibrational symmetry. Column 8 gives the difference between the measured probe-transition wavenumber and the ExoMol prediction. Column 9 gives the Einstein A coefficient predicted by ExoMol. Column 10 gives the difference between the measured probe-transition wavenumber and the Hamiltonian prediction. Column 11 gives the Einstein A coefficient predicted by the Hamiltonian. An asterisk (*) marks a line observed in previous work[3] with lower precision. A double dagger (‡) denotes the same transitions measured in both the $2F_2$ and $3F_2$ AOTR measurements where the same $\nu_3$ Q(2, $F_2$) pump transition was used, for which the weighted mean probe transition wavenumbers and final state term values, together with their corresponding weighted uncertainties, are provided.

| 1 | 2 | 3 | 4 | 5 | 6 | 7 | 8 | 9 | 10 | 11 |
|---|---|---|---|---|---|---|---|---|---|---|
| Pump transition and term value in $P2$ [cm$^{-1}$] | Probe transition wavenumber [cm$^{-1}$] | Final state term value [cm$^{-1}$] | Probe transition integrated absorption coefficient [$10^{-9}$ cm$^{-2}$] | Probe transition width [MHz] | Upper state assignment ($P\ J\ C\ N$) **ExoMol** | Vibrational assignment ($n_1\ n_2\ n_3\ n_4\ C_v$) **ExoMol** | Meas.- **ExoMol** Pred. wavenumber [cm$^{-1}$] | **ExoMol** Einstein A coefficient [s$^{-1}$] | Meas.- **Hamiltonian** Pred. wavenumber [cm$^{-1}$] | **Hamiltonian** Einstein A coefficient [cm$^{-1}$] |
| Measurement: $2F_2$ AOTR and $3F_2$ AOTR<br><br>Pump1: $\nu_3$ Q(2, $F_2$) 3018.65020715(7)[9]<br><br>Term value of $2F_1$ in $P2$: 3050.09259653(7) | 5874.60179(3) | 8924.69439(3) | 0.06(1) | 4.1(8) | 6 3 $F_2$ 481 | 0 0 3 0 $E$ | -0.202120 | 0.007779 | 0.167040 | 0.005329 |
| | 5875.29588(1)‡ | 8925.38848(1) | 0.16(1) | 5.0(5) | 6 3 $F_2$ 482 (483) | 0 0 3 0 $E$ | -0.032110 | 0.019988 | -0.171830 | 0.01079 |
| | 5877.110146(7)‡ | 8927.202742(7) | 0.36(1) | 5.0(2) | 6 2 $F_2$ 375 | 0 5 0 1 $F_2$ | -0.446680 | 0.035295 | 0.205660 | 0.03371 |
| | 5887.24807(3) | 8937.34067(3) | 0.063(9) | 5(1) | 6 2 $F_2$ 378 | 0 0 3 0 $F_2$ | -0.001897 | 0.007006 | -0.002769 | 0.00606 |
| | 5889.441794(2)‡ | 8939.534391(2) | 0.89(1) | 4.63(8) | 6 2 $F_2$ 379 | 0 0 3 0 $F_2$ | -0.032905 | 0.126190 | 0.551410 | 0.01056 |
| | 5889.650778(5)‡ | 8939.743374(5) | 0.295(7) | 4.6(1) | 6 2 $F_2$ 380 | 0 5 0 1 $F_1$ | -1.374100 | 0.000679 | 0.457560 | 0.1003 |
| | 5899.08289(3) | 8949.17549(3) | 0.027(4) | 3.9(9) | 6 3 $F_2$ 499 | 0 5 0 1 $F_2$ | -0.629530 | 0.003133 | -0.248790 | 0.004057 |
| | 5908.650763(2)‡ | 8958.743359(2) | 2.08(1) | 4.40(4) | 6 1 $F_2$ 220 | 0 0 3 0 $F_1$ | -1.110600 | 1.151400 | -0.002800 | 1.127 |
| | 5909.727777(5)‡ | 8959.820374(5) | 0.251(7) | 4.4(2) | 6 3 $F_2$ 506 | 0 5 0 1 $F_2$ | -0.498510 | 0.038040 | 0.318880 | 0.03409 |
| | 5910.59181(3) | 8960.68441(3) | 0.039(6) | 5(1) | 6 3 $F_2$ 508 (507) | 0 3 1 1 $F_1$ | -0.742480 | 0.002341 | 0.277060 | 0.004825 |
| | 5912.25141(3) | 8962.34401(3) | 0.039(7) | 4(1) | 6 3 $F_2$ 509 | 0 5 0 1 $F_2$ | -0.604050 | 0.006799 | 0.043195 | 0.007079 |
| | 5919.248208(4)‡ | 8969.340805(4) | 0.337(5) | 4.6(1) | 6 3 $F_2$ 512 | 0 0 3 0 $F_2$ | -0.034907 | 0.045475 | 0.253540 | 0.04023 |
| | 5921.749057(2)‡ | 8971.841654(2) | 0.608(5) | 4.67(6) | 6 3 $F_2$ 513 | 0 0 3 0 $F_2$ | -0.000638 | 0.083443 | 0.192840 | 0.07169 |
| | 5923.496498(5)‡ | 8973.589094(5) | 0.268(6) | 4.7(1) | 6 2 $F_2$ 384 | 0 2 2 0 $A_1$ | -0.432070 | 0.034554 | -0.466900 | 0.0233 |
| | 5925.99692(3) | 8976.08952(3) | 0.056(6) | 6(1) | 6 2 $F_2$ 385 | 0 5 0 1 $F_1$ | -1.043700 | 0.003069 | -0.016003 | 0.009223 |
| | 5928.611418(2)*‡ | 8978.704014(2) | 1.47(1) | 4.86(5) | 6 2 $F_2$ 386 | 0 0 3 0 $F_1$ | -1.084000 | 0.147570 | -0.062078 | 0.1563 |
| | 5944.406096(6)*‡ | 8994.498693(6) | 0.260(7) | 4.5(2) | 6 2 $F_2$ 391 | 0 5 0 1 $F_1$ | -0.795940 | 0.027893 | -0.194430 | 0.0297 |
| | 5958.673581(2)*‡ | 9008.766177(2) | 3.91(2) | 4.77(4) | 6 3 $F_2$ 520 | 0 0 3 0 $F_1$ | -1.094500 | 0.515390 | -0.040073 | 0.5039 |
| | 5959.386802(2)*‡ | 9009.479398(2) | 2.36(2) | 4.94(5) | 6 2 $F_2$ 393 | 1 0 2 0 $A_1$ | -0.472830 | 0.237740 | 0.087081 | 0.2387 |
| | 5990.090523(2)‡ | 9040.183119(2) | 1.64(2) | 4.86(9) | 6 2 $F_2$ 398 | 0 2 2 0 $A_1$ | -0.110470 | 0.171360 | -0.008439 | 0.1517 |
| | 6006.75156(2) | 9056.84416(2) | 0.09(1) | 3.3(7) | 6 1 $F_2$ 228 | 0 0 3 0 $F_2$ | 0.000094 | 0.045566 | -0.000278 | 0.04424 |
| | 6025.548408(3)‡ | 9075.641004(3) | 1.50(3) | 4.8(1) | 6 2 $F_2$ 403 | 0 0 3 0 $F_2$ | 0.001043 | 0.162910 | -0.006676 | 0.1588 |
| | 6055.86878(1) | 9105.96137(1) | 1.55(8) | 5.3(4) | 6 3 $F_2$ 544 | 0 0 3 0 $F_2$ | -0.024632 | 0.189170 | -0.031075 | 0.1796 |
| | 6057.46411(2) | 9107.55670(2) | 0.25(4) | 3.6(7) | 6 3 $F_2$ 545 | 0 0 3 0 $F_2$ | -0.000359 | 0.047432 | -0.002134 | 0.04815 |
| | 5880.22470(5) | 8959.82041(5) | 0.057(9) | 7(1) | 6 3 $F_2$ 506 | 0 5 0 1 $F_2$ | -0.498480 | 0.018019 | 0.319650 | 0.0169 |
| Measurement:<br>4$F_2$ AOTR<br><br>Pump1:<br>$\nu_3$ R(2, $F_2$)<br>3048.15331811(8)[10]<br><br>Term value of 3$F_1$<br>in $P2$:<br>3079.59570750(9) | 5884.40826(4) | 8964.00397(4) | 0.08(1) | 6(1) | 6 4 $F_2$ 629 (630) | 0 3 1 1 $A_2$ | 0.412150 | 0.003899 | -0.306000 | 0.006746 |
| | 5889.74512(3) | 8969.34083(3) | 0.08(1) | 5.8(9) | 6 3 $F_2$ 512 | 0 0 3 0 $F_2$ | -0.034880 | 0.025708 | 0.254300 | 0.02239 |
| | 5892.24598(3) | 8971.84169(3) | 0.11(1) | 6.5(8) | 6 3 $F_2$ 513 | 0 0 3 0 $F_2$ | -0.000604 | 0.029456 | 0.193610 | 0.02467 |
| | 5892.90050(5) | 8972.49621(5) | 0.06(1) | 6(1) | 6 4 $F_2$ 637 (638) | 0 1 2 1 $F_2$ | 0.684230 | 0.016776 | -0.399350 | 0.002853 |
| | 5893.993379(4) | 8973.589087(4) | 1.01(1) | 5.83(9) | 6 2 $F_2$ 384 | 0 2 2 0 $A_1$ | -0.43208 | 0.13523 | -0.466170 | 0.1124 |
| | 5896.493854(5) | 8976.089562(5) | 0.68(1) | 5.7(1) | 6 2 $F_2$ 385 | 0 5 0 1 $F_1$ | -1.043700 | 0.054610 | -0.015228 | 0.1426 |
| | 5899.108299(2) | 8978.704006(2) | 4.91(3) | 5.96(5) | 6 2 $F_2$ 386 | 0 0 3 0 $F_1$ | -1.084000 | 0.609010 | -0.000221 | 0.5442 |
| | 5913.18730(2) | 8992.78301(2) | 0.14(1) | 5.2(5) | 6 4 $F_2$ 656 | 0 5 0 1 $F_2$ | -0.624620 | 0.011882 | -0.325980 | 0.01134 |
| | 5918.14138(2) | 8997.73709(2) | 0.21(1) | 6.4(5) | 6 4 $F_2$ 661 | 0 5 0 1 $F_2$ | -0.503250 | 0.014277 | -0.000236 | 0.01725 |
| | 5920.77557(2) | 9000.37128(2) | 0.046(9) | 2.6(7) | 6 4 $F_2$ 662 (663) | 0 5 0 1 $F_1$ | 0.034711 | 0.002647 | -0.000140 | 0.01023 |
| | 5923.32500(4) | 9002.92071(4) | 0.12(2) | 6(1) | 6 4 $F_2$ 665 (664) | 0 3 1 1 $F_1$ | -0.517930 | 0.003961 | -0.000237 | 0.00147 |
| | 5923.948504(5) | 9003.544212(5) | 0.81(1) | 6.0(1) | 6 4 $F_2$ 666 (665) | 0 5 0 1 $F_2$ | -0.420470 | 0.080321 | -0.000176 | 0.005658 |
| | 5924.26540(2) | 9003.86111(2) | 0.41(3) | 6.7(5) | 6 4 $F_2$ 667 (666) | 0 5 0 1 $F_2$ | -0.560490 | 0.032439 | -0.000168 | 0.08091 |
| | 5924.99664(3) | 9004.59235(3) | 0.09(1) | 4.6(8) | 6 4 $F_2$ 669 (667) | 0 3 1 1 $F_1$ | -0.532420 | 0.002285 | -0.000160 | 0.002162 |
| | 5929.170464(2) | 9008.766172(2) | 3.60(2) | 5.75(5) | 6 3 $F_2$ 520 | 0 0 3 0 $F_1$ | -1.094500 | 1.009700 | -0.000205 | 0.9923 |
| | 5929.883686(3) | 9009.479394(3) | 1.97(2) | 5.96(7) | 6 2 $F_2$ 393 | 1 0 2 0 $A_1$ | -0.472830 | 0.230290 | -0.000206 | 0.233 |
| | 5932.27923(1) | 9011.87493(1) | 0.22(1) | 6.1(5) | 6 4 $F_2$ 675 | 0 0 3 0 $F_2$ | -0.001557 | 0.019196 | -0.000165 | 0.01671 |
| | 5935.245189(3) | 9014.840896(3) | 1.72(2) | 6.03(8) | 6 4 $F_2$ 677 | 0 0 3 0 $F_2$ | -0.106850 | 0.134070 | -0.000212 | 0.1149 |
| | 5960.587406(3) | 9040.183113(3) | 1.62(3) | 6.0(1) | 6 2 $F_2$ 398 | 0 2 2 0 $A_1$ | -0.110470 | 0.192310 | -0.000184 | 0.1723 |

**Table S4. List of all V-type $P4 \leftarrow P0$ transitions from the $2F_2$ ground state detected in our work.** Column 1 lists the measured center wavenumbers, with the uncertainties given in parentheses. Column 2 lists the upper state assignments (*P J C N*) based on ExoMol predictions, where *P* is the polyad number, *J* the total angular momentum, *C* the rovibrational symmetry, and *N* the counting number. Column 3 lists the vibrational assignments ($n_1$ $n_2$ $n_3$ $n_4$ $C_v$) from ExoMol, where $n_1$, $n_2$, $n_3$, $n_4$ are vibrational modes, and $C_v$ the vibrational symmetry. Column 4 provides the differences between the measured probe transition wavenumbers and the ExoMol predictions. Column 5 shows the predicted Einstein A coefficients from ExoMol. Column 6 provides the differences between the measured probe transition wavenumbers and the Hamiltonian predictions. Column 7 shows the predicted Einstein A coefficients from the Hamiltonian. Column 8 provides the differences between the measured probe transition wavenumbers and the HITRAN line list. Column 9 shows the predicted line intensity from HITRAN. Column 10 provides the differences between the measured probe transition wavenumbers and the WKLMC line list. Column 11 shows the predicted line intensity from the WKLMC line list. The line marked by an asterisk (*) has been observed in previous work[8] with worse precision.

| 1 | 2 | 3 | 4 | 5 | 6 | 7 | 8 | 9 | 10 | 11 |
|---|---|---|---|---|---|---|---|---|---|---|
| Probe transition wavenumber [$cm^{-1}$] | Upper state assignment (*P*, *J*, *C*, *N*) **ExoMol** | Vibrational assignment ($n_1$ $n_2$ $n_3$ $n_4$ $C_v$) **ExoMol** | Meas.- **ExoMol** Pred. wavenumber [$10^{-6}$ $cm^{-1}$] | **ExoMol** Einstein A coefficient [$s^{-1}$] | Meas.- **Hamiltonian** Pred. wavenumber [$10^{-6}$ $cm^{-1}$] | **Hamiltonian** Einstein A coefficient [$s^{-1}$] | Meas.- **HITRAN** Pred. wavenumber [$10^{-6}$ $cm^{-1}$] | **HITRAN** line intensity [$10^{-24}$ cm/molec] | Meas.- **WKLMC** Pred. wavenumber [$10^{-6}$ $cm^{-1}$] | **WKLMC** line intensity [$10^{-24}$ cm/molec] |
| 5838.35795(3) | 4 3 $F_1$ 77 | 0 1 1 1 $F_2$ | 10.71 | 0.026295 | -1370.3 | 0.02641 | -26.29 | 34.09 | - | - |
| 5840.588601(9) | 4 3 $F_1$ 78 | 0 1 1 1 $F_2$ | -454.28 | 0.084362 | -1241.3 | 0.08382 | -488.28 | 102.90 | - | - |
| 5845.99483(2) | 4 2 $F_1$ 57 | 0 1 1 1 $A_2$ | -1158.70 | 0.026257 | 35.35 | 0.02716 | -674.65 | 25.34 | - | - |
| 5846.309198(7) | 4 3 $F_1$ 79 | 0 1 1 1 $A_1$ | 3.22 | 0.060144 | 583.22 | 0.06266 | -771.78 | 75.74 | - | - |
| 5847.60270(2) | 4 2 $F_1$ 58 | 0 1 1 1 $F_1$ | -541.16 | 0.014196 | -1845.2 | 0.01428 | -462.16 | 12.78 | - | - |
| 5848.01514(2) | 4 3 $F_1$ 80 | 0 1 1 1 *E* | -130.98 | 0.015114 | -658.98 | 0.01606 | -594.98 | 20.60 | - | - |
| 5854.193311(7) | 4 3 $F_1$ 81 | 0 1 1 1 $F_1$ | -16.84 | 0.044896 | 175.16 | 0.0462 | 14.16 | 58.16 | - | - |
| 5863.74788(2) | 4 3 $F_1$ 84 | 0 1 1 1 $F_1$ | -252.33 | 0.0096998 | 40.671 | 0.01013 | -411.33 | 12.16 | -11.33 | 12.08 |
| 5865.11898(2) | 4 2 $F_1$ 63 | 0 3 0 1 $F_2$ | -241.70 | 0.0074209 | -55.699 | 0.005938 | -437.70 | 7.95 | -47.70 | 7.29 |
| 5878.165093(6) | 4 3 $F_1$ 87 | 0 1 1 1 $F_2$ | 288.15 | 0.024272 | 1158.1 | 0.02552 | -408.85 | 39.88 | -308.85 | 33.02 |
| 5883.700719(9) | 4 3 $F_1$ 89 | 0 1 1 1 *E* | 143.73 | 0.0085649 | 270.73 | 0.008816 | -556.27 | 11.26 | -376.27 | 11.09 |
| 5983.19366(1) | 4 1 $F_1$ 51 | 0 0 2 0 $F_2$ | -6.72 | 0.56568 | 371.28 | 0.5599 | -1.72 | 274.00 | 357.28 | 277.90 |
| 6004.64357(1)* | 4 2 $F_1$ 73 | 0 0 2 0 $F_2$ | -28.41 | 0.60776 | 256.59 | 0.6021 | -37.41 | 487.00 | 534.59 | 469.30 |
| 6054.81735(2) | 4 2 $F_1$ 75 | 0 2 1 0 $F_2$ | -292.10 | 0.034606 | -253.1 | 0.03377 | -47.10 | 27.90 | -251.10 | 28.05 |

**Table S5. List of all final state term values in the *P*8, *P*6 and *P*4 polyads assigned to ExoMol.** Column 1 lists the experimental final state term values, with the uncertainties given in parentheses. Column 2 lists the upper state assignments (*P J C N*) based on ExoMol predictions, where *P* is the polyad number, *J* the total angular momentum, *C* the rovibrational symmetry, and *N* the counting number. Column 3 lists the vibrational assignments ($n_1$ $n_2$ $n_3$ $n_4$ $C_v$) from ExoMol, where $n_1$, $n_2$, $n_3$, $n_4$ are vibrational modes, and $C_v$ the vibrational symmetry. Column 4 shows the predicted Einstein A coefficients from ExoMol. Column 5 provides the predicted final state term values from ExoMol. Column 6 indicates for which pump transitions the levels were reached.

| 1 | 2 | 3 | 4 | 5 | 6 |
|---|---|---|---|---|---|
| Measured final state term values [$cm^{-1}$] | Upper state assignment (*P J C N*) **ExoMol** | Vibrational assignment ($n_1$ $n_2$ $n_3$ $n_4$ $C_v$) **ExoMol** | Einstein A coefficient [$s^{-1}$] **ExoMol** | **ExoMol** Pred. Final state term value [$cm^{-1}$] | Observed via pump transitions |
| 12116.18957(2) | 8 3 $F_1$ 2063 | 0 4 2 0 $A_1$ | 0.18193 | 12117.56965 | Q(2, $F_2$) then R(2, $F_1$) |
| 12077.42399(2) | 8 5 $F_1$ 3093 | 0 5 1 1 $F_2$ | 0.027941 | 12080.22707 | R(2, $F_2$) then R(3, $F_1$) |
| 12062.204589(8) | 8 4 $F_1$ 2532 | 0 0 4 0 $E$ | 0.11239 | 12064.31975 | Q(2, $F_2$) then R(2, $F_1$) |
| 12061.00467(1) | 8 4 $F_1$ 2531 | 0 0 4 0 $E$ | 0.15157 | 12063.11276 | R(2, $F_2$) then R(3, $F_1$) |
| 12039.802930(9) | 8 4 $F_1$ 2513 | 0 0 4 0 $F_2$ | 0.033202 | 12042.85816 | R(2, $F_2$) then R(3, $F_1$) |
| 12029.399561(8) | 8 4 $F_1$ 2499 | 0 5 1 1 $F_2$ | 0.010711 | 12031.18641 | Q(2, $F_2$) then R(2, $F_1$), R(2, $F_2$) then R(3, $F_1$) |
| 12026.986150(4) | 8 4 $F_1$ 2497 | 0 5 1 1 $F_2$ | 0.10658 | 12029.99013 | Q(2, $F_2$) then R(2, $F_1$), R(2, $F_2$) then R(3, $F_1$) |
| 12021.356736(3) | 8 3 $F_1$ 2011 | 0 0 4 0 $E$ | 0.22453 | 12023.34664 | Q(2, $F_2$) then Q(2, $F_1$), Q(2, $F_2$) then R(2, $F_1$), R(2, $F_2$) then R(3, $F_1$) |
| 12020.080117(6) | 8 4 $F_1$ 2489 | 0 0 4 0 $F_2$ | 0.13519 | 12023.17513 | Q(2, $F_2$) then R(2, $F_1$) |
| 11995.44591(2) | 8 3 $F_1$ 1998 | 0 0 4 0 $F_2$ | 0.0065005 | 11998.50781 | Q(2, $F_2$) then R(2, $F_1$) |
| 11992.49140(1) | 8 4 $F_1$ 2459 | 1 4 1 0 $F_2$ | 0.035685 | 11995.7811 | Q(2, $F_2$) then R(2, $F_1$) |
| 11992.02236(1) | 8 2 $F_1$ 1399 | 0 0 4 0 $E$ | 0.015362 | 11993.98737 | Q(2, $F_2$) then Q(2, $F_1$), Q(2, $F_2$) then R(2, $F_1$) |
| 11991.71457(1) | 8 4 $F_1$ 2458 | 1 4 1 0 $F_2$ | 0.058819 | 11995.13523 | Q(2, $F_2$) then R(2, $F_1$) |
| 11983.107008(3) | 8 3 $F_1$ 1986 | 0 0 4 0 $F_2$ | 0.012713 | 11986.10948 | Q(2, $F_2$) then R(2, $F_1$), R(2, $F_2$) then R(3, $F_1$) |
| 11982.706266(3) | 8 3 $F_1$ 1985 | 0 5 1 1 $F_2$ | 0.038364 | 11985.21555 | Q(2, $F_2$) then Q(2, $F_1$), Q(2, $F_2$) then R(2, $F_1$), R(2, $F_2$) then R(3, $F_1$) |
| 11981.16703(2) | 8 3 $F_1$ 1984 | 0 0 4 0 $F_2$ | 0.056958 | 11984.21392 | R(2, $F_2$) then R(3, $F_1$) |
| 11971.656999(6) | 8 1 $F_1$ 877 | 0 0 4 0 $E$ | 0.33993 | 11973.61388 | Q(2, $F_2$) then Q(2, $F_1$) |
| 11964.599897(5) | 8 2 $F_1$ 1391 | 0 2 3 0 $F_1$ | 0.021224 | 11967.00929 | Q(2, $F_2$) then Q(2, $F_1$), Q(2, $F_2$) then R(2, $F_1$) |
| 11961.108450(8) | 8 2 $F_1$ 1388 | 0 0 4 0 $F_2$ | 0.090837 | 11963.69141 | Q(2, $F_2$) then Q(2, $F_1$), Q(2, $F_2$) then R(2, $F_1$) |
| 11959.795847(8) | 8 2 $F_1$ 1386 | 0 0 4 0 $F_2$ | 0.021886 | 11962.58762 | Q(2, $F_2$) then Q(2, $F_1$), Q(2, $F_2$) then R(2, $F_1$) |
| 11952.873184(4) | 8 2 $F_1$ 1383 | 0 5 1 1 $F_2$ | 0.066754 | 11954.88522 | Q(2, $F_2$) then Q(2, $F_1$), Q(2, $F_2$) then R(2, $F_1$) |
| 11950.337025(4) | 8 2 $F_1$ 1381 | 0 5 1 1 $F_2$ | 0.066386 | 11952.86277 | Q(2, $F_2$) then Q(2, $F_1$), Q(2, $F_2$) then R(2, $F_1$) |
| 11948.424812(8) | 8 3 $F_1$ 1956 | 1 4 1 0 $F_2$ | 0.079881 | 11951.88322 | Q(2, $F_2$) then Q(2, $F_1$), Q(2, $F_2$) then R(2, $F_1$), R(2, $F_2$) then R(3, $F_1$) |
| 11914.491122(7) | 8 2 $F_1$ 1358 | 1 4 1 0 $F_2$ | 0.15939 | 11918.16791 | Q(2, $F_2$) then Q(2, $F_1$) |
| 9107.55670(2) | 6 3 $F_2$ 545 | 0 0 3 0 $F_2$ | 0.047432 | 9107.557038 | Q(2, $F_2$) |
| 9105.96137(1) | 6 3 $F_2$ 544 | 0 0 3 0 $F_2$ | 0.18917 | 9105.985982 | Q(2, $F_2$) |
| 9075.641004(3) | 6 2 $F_2$ 403 | 0 0 3 0 $F_2$ | 0.16291 | 9075.639938 | Q(2, $F_2$) |
| 9056.84416(2) | 6 1 $F_2$ 228 | 0 0 3 0 $F_2$ | 0.045566 | 9056.844041 | Q(2, $F_2$) |
| 9040.183117(2) | 6 2 $F_2$ 398 | 0 2 2 0 $A_1$ | 0.17136 | 9040.293562 | Q(2, $F_2$), R(2, $F_2$) |
| 9014.840896(3) | 6 4 $F_2$ 677 | 0 0 3 0 $F_2$ | 0.13407 | 9014.947722 | R(2, $F_2$) |
| 9011.87493(1) | 6 4 $F_2$ 675 | 0 0 3 0 $F_2$ | 0.019196 | 9011.876468 | R(2, $F_2$) |
| 9009.479397(2) | 6 2 $F_2$ 393 | 1 0 2 0 $A_1$ | 0.23774 | 9009.952201 | Q(2, $F_2$), R(2, $F_2$) |
| 9008.766175(1) | 6 3 $F_2$ 520 | 0 0 3 0 $F_1$ | 0.51539 | 9009.860622 | Q(2, $F_2$), R(2, $F_2$) |
| 9004.59235(3) | 6 4 $F_2$ 669 | 0 3 1 1 $F_1$ | 0.0022851 | 9005.124741 | R(2, $F_2$) |
| 9003.86111(2) | 6 4 $F_2$ 667 | 0 5 0 1 $F_2$ | 0.032439 | 9004.421578 | R(2, $F_2$) |
| 9003.544212(5) | 6 4 $F_2$ 666 | 0 5 0 1 $F_2$ | 0.080321 | 9003.964654 | R(2, $F_2$) |
| 9002.92071(4) | 6 4 $F_2$ 665 | 0 3 1 1 $F_1$ | 0.0039611 | 9003.438619 | R(2, $F_2$) |
| 9000.37128(2) | 6 4 $F_2$ 662 | 0 5 0 1 $F_1$ | 0.0026465 | 9000.336546 | R(2, $F_2$) |
| 8997.73709(2) | 6 4 $F_2$ 661 | 0 5 0 1 $F_2$ | 0.014277 | 8998.240319 | R(2, $F_2$) |
| 8994.498693(6) | 6 2 $F_2$ 391 | 0 5 0 1 $F_1$ | 0.027893 | 8995.294612 | Q(2, $F_2$) |
| 8992.78301(2) | 6 4 $F_2$ 656 | 0 5 0 1 $F_2$ | 0.011882 | 8993.407603 | R(2, $F_2$) |
| 8978.704011(2) | 6 2 $F_2$ 386 | 0 0 3 0 $F_1$ | 0.14757 | 8979.78798 | Q(2, $F_2$), R(2, $F_2$) |
| 8976.089561(5) | 6 2 $F_2$ 385 | 0 5 0 1 $F_1$ | 0.0030689 | 8977.133219 | Q(2, $F_2$), R(2, $F_2$) |
| 8973.589089(3) | 6 2 $F_2$ 384 | 0 2 2 0 $A_1$ | 0.034554 | 8974.021143 | Q(2, $F_2$), R(2, $F_2$) |
| 8971.841654(2) | 6 3 $F_2$ 513 | 0 0 3 0 $F_2$ | 0.083443 | 8971.842268 | Q(2, $F_2$), R(2, $F_2$) |
| 8972.49621(5) | 6 4 $F_2$ 637 | 0 1 2 1 $F_2$ | 0.016776 | 8971.811951 | R(2, $F_2$) |
| 8969.340805(4) | 6 3 $F_2$ 512 | 0 0 3 0 $F_2$ | 0.045475 | 8969.375688 | Q(2, $F_2$), R(2, $F_2$) |
| 8964.00397(4) | 6 4 $F_2$ 629 | 0 3 1 1 $A_2$ | 0.0038993 | 8963.591795 | R(2, $F_2$) |
| 8962.34401(3) | 6 3 $F_2$ 509 | 0 5 0 1 $F_2$ | 0.0067991 | 8962.948041 | Q(2, $F_2$) |
| 8960.68441(3) | 6 3 $F_2$ 508 | 0 3 1 1 $F_1$ | 0.0023406 | 8961.426867 | Q(2, $F_2$) |
| 8959.820374(6) | 6 3 $F_2$ 506 | 0 5 0 1 $F_2$ | 0.03804 | 8960.318863 | Q(2, $F_2$), R(2, $F_2$) |
| 8958.743359(2) | 6 1 $F_2$ 220 | 0 0 3 0 $F_1$ | 1.1514 | 8959.853898 | Q(2, $F_2$) |
| 8949.17549(3) | 6 3 $F_2$ 499 | 0 5 0 1 $F_2$ | 0.0031331 | 8949.804998 | Q(2, $F_2$) |
| 8939.743374(5) | 6 2 $F_2$ 380 | 0 5 0 1 $F_1$ | 0.00067864 | 8941.117493 | Q(2, $F_2$) |
| 8939.534391(2) | 6 2 $F_2$ 379 | 0 0 3 0 $F_2$ | 0.12619 | 8939.567272 | Q(2, $F_2$) |
| 8937.34067(3) | 6 2 $F_2$ 378 | 0 0 3 0 $F_2$ | 0.0070061 | 8937.342541 | Q(2, $F_2$) |
| 8927.202742(7) | 6 2 $F_2$ 375 | 0 5 0 1 $F_2$ | 0.035295 | 8927.649402 | Q(2, $F_2$) |
| 8925.38848(1) | 6 3 $F_2$ 482 | 0 0 3 0 $E$ | 0.019988 | 8925.420568 | Q(2, $F_2$) |
| 8924.69439(3) | 6 3 $F_2$ 481 | 0 0 3 0 $E$ | 0.007779 | 8924.896484 | Q(2, $F_2$) |
| 6086.25974(2) | 4 2 $F_1$ 75 | 0 2 1 0 $F_2$ | 0.034606 | 6086.260007 | Q(2, $F_2$) |
| 6036.08596(1) | 4 2 $F_1$ 73 | 0 0 2 0 $F_2$ | 0.60776 | 6036.085929 | Q(2, $F_2$), R(2, $F_2$) |
| 6014.636047(7) | 4 1 $F_1$ 51 | 0 0 2 0 $F_2$ | 0.56568 | 6014.63603 | Q(2, $F_2$), R(2, $F_2$) |
| 5915.14311(2) | 4 3 $F_1$ 89 | 0 1 1 1 $E$ | 0.0085649 | 5915.142946 | Q(2, $F_2$), R(2, $F_2$) |
| 5909.607483(8) | 4 3 $F_1$ 87 | 0 1 1 1 $F_2$ | 0.024272 | 5909.607169 | Q(2, $F_2$), R(2, $F_2$) |
| 5896.56137(2) | 4 2 $F_1$ 63 | 0 3 0 1 $F_2$ | 0.0074209 | 5896.56156 | Q(2, $F_2$), R(2, $F_2$) |
| 5895.19027(2) | 4 3 $F_1$ 84 | 0 1 1 1 $F_1$ | 0.0096998 | 5895.190507 | Q(2, $F_2$), R(2, $F_2$) |
| 5885.63570(1) | 4 3 $F_1$ 81 | 0 1 1 1 $F_1$ | 0.044896 | 5885.635687 | Q(2, $F_2$), R(2, $F_2$) |
| 5879.45752(2) | 4 3 $F_1$ 80 | 0 1 1 1 $E$ | 0.015114 | 5879.457632 | R(2, $F_2$) |
| 5879.04509(2) | 4 2 $F_1$ 58 | 0 1 1 1 $F_1$ | 0.014196 | 5879.045605 | R(2, $F_2$) |
| 5877.75159(2) | 4 3 $F_1$ 79 | 0 1 1 1 $A_1$ | 0.060144 | 5877.751561 | Q(2, $F_2$), R(2, $F_2$) |
| 5877.43721(2) | 4 2 $F_1$ 57 | 0 1 1 1 $A_2$ | 0.026257 | 5877.43835 | R(2, $F_2$) |
| 5872.03099(2) | 4 3 $F_1$ 78 | 0 1 1 1 $F_2$ | 0.084362 | 5872.031422 | Q(2, $F_2$), R(2, $F_2$) |
| 5869.80034(3) | 4 3 $F_1$ 77 | 0 1 1 1 $F_2$ | 0.026295 | 5869.800309 | R(2, $F_2$) |